\documentclass[%
reprint,
amsmath,amssymb,
aps,
prx,
floatfix,
]{revtex4-2}
\usepackage{graphicx}% Include figure files
\usepackage{dcolumn}% Align table columns on decimal point
\usepackage{bm}% bold math
\usepackage{color}
\definecolor{darkred}{rgb}{0.6,0,0}
\usepackage{hyperref}
\newcommand\bfr{{\bf r}}
\newcommand\bfu{{\bf u}}
\newcommand\bfJ{{\bf J}}
\newcommand\bfcJ{{\boldsymbol{\mathcal{J}}}}
\newcommand\bfm{{\bf m}}

\usepackage[normalem]{ulem}

\begin{document}
	
	\preprint{APS/123-QED}
	
	\title{Disordered boundaries destroy bulk phase separation in scalar active matter}
	
	\author{Ydan Ben Dor}
	\thanks{These two authors contributed equally.}
	\author{Sunghan Ro}
	\thanks{These two authors contributed equally.}
	\author{Yariv Kafri}
	\affiliation{Department of Physics, Technion -- Israel Institute of Technology, Haifa 32000, Israel}
	
	\author{Mehran Kardar}
	\affiliation{Department of Physics, Massachusetts Institute of Technology, Cambridge, Massachusetts 02139, USA}
	
	\author{Julien Tailleur}
	\affiliation{Universit\'{e} de Paris, Laboratoire Mati\`{e}re et Syst\`{e}mes Complexes (MSC), UMR 7057 CNRS, F-75205 Paris, France}
	
	\begin{abstract}
		We show that disordered boundaries destroy bulk phase separation in scalar active systems in dimension $d<d_c=3$. This is in strong contrast with the equilibrium case where boundaries have no impact on the bulk of phase-separated systems. The underlying mechanism is revealed by considering a localized deformation of an otherwise flat wall, from which the case of a disordered boundary can be inferred. We find long-ranged correlations of the density field as well as a cascade of eddies which we show prevent bulk phase separation in low enough dimensions. The results are derived for dilute systems as well as in the presence of interactions, under the sole condition that the density field is the unique hydrodynamic mode. Our theoretical calculations are validated by numerical simulations of microscopic active systems.
	\end{abstract}
	
	\maketitle
	
	\section{Introduction}
	Active matter refers to a class of non-equilibrium systems in which individual particles are self-propelled due to an irreversible consumption of energy. Their physics is relevant to systems ranging from biological to man-made materials~\cite{vicsek2012collective,romanczuk2012active,marchetti_hydrodynamics_2013,cates_motility-induced_2015,Bechinger2016RMP,gnesotto2018broken,o2021time}.
	They have attracted much attention since they exhibit a host of novel collective behaviors which cannot be found in equilibrium systems. Examples range from the transition to collective motion, through low-Reynolds turbulence, to motility-induced phase separation (MIPS)~\cite{toner_flocks_1998,tailleur_statistical_2008,thompson_lattice_2011,fily_athermal_2012,wensink2012meso,palacci_living_2013,buttinoni_dynamical_2013,cates_when_2013,stenhammar_continuum_2013,redner_structure_2013,solon_active_2015,redner_classical_2016,paliwal_chemical_2018,cates_motility-induced_2015,solon_generalized_2018,tjhung_cluster_2018,kourbane-houssene_exact_2018,whitelam_phase_2018,geyer_freezing_2019,chate2020dry,o2021time}. The latter corresponds to the ability of active systems to phase separate, even when there are no attractive interactions between the particles.

	\begin{figure*}[h!]
		\includegraphics[width=1.0\linewidth]{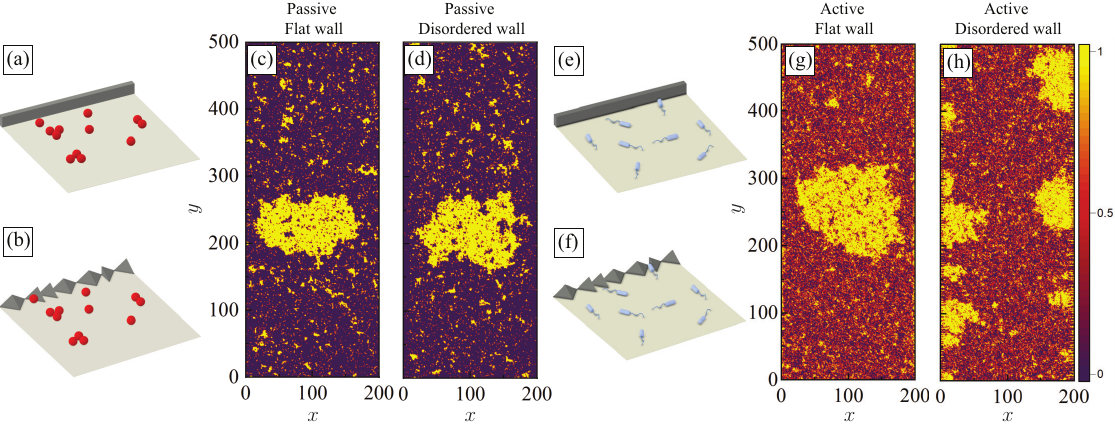}
		\caption{Impact of flat (a \& e) and disordered walls (b \& f) on phase separation in passive  (a \& b) and active (e \& f) systems. In the presence of attractive interactions, simulations of a passive lattice gas at low temperature shows phase separation in both settings (panels c and  d). In contrast, simulations of an active lattice gas (panel h) show that the disordered boundary destroys the phase separation observed in the presence of a flat wall (panel g). Color encodes density; see Appendix~\ref{app:simulation details} for further numerical details.
		}\label{fig:schematic}
	\end{figure*}

	It has been long realized,  experimentally and theoretically~\cite{kudrolli2008swarming,Deseigne2010PRL,woodhouse2012spontaneous,wioland2013confinement,bricard2015emergent,Bechinger2016RMP,wioland2016ferromagnetic,souslov2017topological} that the {\it shapes} of boundaries in active systems lead to interesting effects, from the rotation of asymmetric gears~\cite{Sokolov2010PNAS,DiLeonardo2010PNAS} to the emergence of ratchet currents~\cite{nikola_active_2016,Bechinger2016RMP}. It is tempting to assume that these effects are localized to the wall, on microscopic scales set by the particles' persistence lengths, the potential shapes, and the correlation lengths set by interactions. Consequently, much of the theoretical work on bulk collective behaviors, in particular for dry scalar active matter, has focused on systems which are either infinite or subject to periodic boundary conditions~\cite{Marchetti2013RMP,cates_motility-induced_2015,chate2019dry}. The underlying salient assumption is that, much like in equilibrium, the precise nature of the boundaries only affects a sub-extensive region in macroscopic active systems and thus does not influence their \textit{bulk} behaviors. 
	
\begin{figure*}[t!]
		\includegraphics[width=1\linewidth]{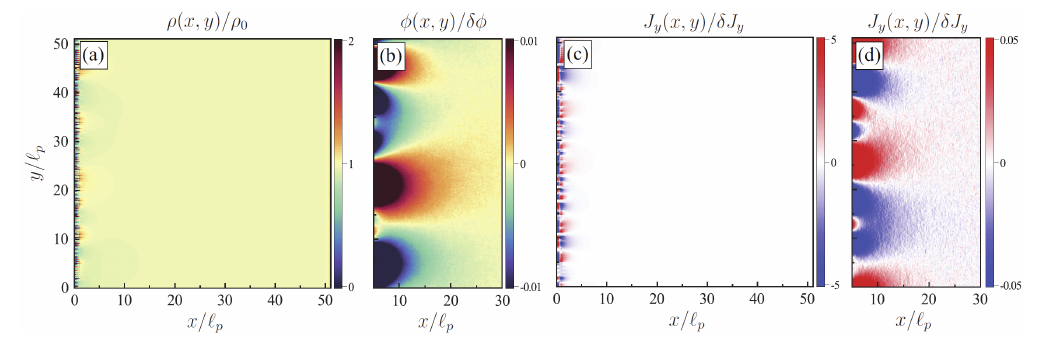}
		\caption{Steady-state density and currents for run-and-tumble particles on a lattice, in the presence of a disordered wall at $x=0$ and periodic boundary conditions along the $\hat {\bf y}$ direction. The disordered wall is modelled as a random potential which vanishes for $x$ larger than the particle persistence length $\ell_p$; for other numerical details, see Appendix~\ref{app:simulation details}.  {\bf (a)} Particle density $\rho(x,y)$ in the full system, normalized by the average density. Lengths are rescaled by the particle run length $\ell_p$. Note the presence of a strong density accumulation close to the wall at $x=0$ that is modulated by the disorder. The color code corresponds to $\rho(x,y)/\rho_0$. 
			{\bf (b)} Density modulation $\phi(x,y) \equiv \rho(x,y) - \langle{\rho(x)}\rangle$ in the bulk of the system, where $\langle{\rho(x)}\rangle$ is the average density at a distance $x$ from the wall, normalized by the standard deviation of the density modulation $\delta \phi$. The density modulations extend deep in the bulk of the system, far beyond the microscopic scales set by the particle run length and the disordered wall. (Note that the $x$ axis starts at $5$ run lengths from the wall.)
			{\bf (c)} Current along the $\hat {\bf y}$ direction in the full system, normalized by the current standard deviation $\delta J_y$. At this scale, a localized current flowing along the wall is observed, as expected from the existing literature~\cite{nikola_active_2016}.
			{\bf (d)} A close-up on the bulk region shown in panel (b) reveals the existence of large eddies whose scales increase with the distance from the wall. 
		}\label{fig:rho_and_jy}
	\end{figure*}

	In this article, we show that this is generically not the case, even for dry, scalar active matter where boundaries are expected to have the weakest influence. This is illustrated in Fig.~\ref{fig:schematic} which compares the fate of passive and active phase separation in the presence of a disordered wall. As expected~\cite{lebowitz_statistical_1999}, the disordered boundary leaves the phase-separated equilibrium system unaffected (Fig.~\ref{fig:schematic}a-d). In striking contrast, the disordered wall washes out phase separation in the active case (Fig.~\ref{fig:schematic}e-h), thus strongly altering the phase diagram. In fact, we demonstrate that phase separation is destroyed by disordered boundaries in dimension $d < d_c=3$. As we show below, this is a result of disordered boundaries inducing scale-free density modulations and eddy cascades deep in the bulk of active systems. These can already be seen, upon close inspection, in the dilute limit, as illustrated in Fig.~\ref{fig:rho_and_jy}, showing that disordered boundaries do not solely lead to the localized effects that had been reported earlier~\cite{elgeti_wall_2013,elgeti2009self,tailleur_sedimentation_2009,nikola_active_2016}.

	 To investigate the physics behind the numerical results reported in Figs.~\ref{fig:schematic} and~\ref{fig:rho_and_jy}, we start, in Sec.~\ref{sec:multipole expansion}, by considering a dilute  system in the presence of a localized deformation on an otherwise flat wall. We show that it induces non-standard boundary conditions on the density and current fields. Using appropriate Green's functions, we show that the perturbation induces a long-range modulation in the steady-state density profile, which we characterize in the far field limit. We then show in Sec.~\ref{sec: disordered boundaries} how these results allow us to describe a disordered wall and to evaluate the disorder-averaged two-point correlation functions of the density and current fields. These results, first derived in the dilute limit in Sec.~\ref{sec: disordered boundaries: dilute systems} are then generalized to interacting systems in Sec.~\ref{sec: disordered boundaries: interacting systems}. Finally, we show in Sec.~\ref{sec:Imry-Ma} that, even though the density modulations and currents decay as power laws in the bulk of the system, they are sufficient to destroy MIPS in dimension $d < d_c=3$. In practice, the wall creates a disordered combination of long-range attractive and repulsive forces that prevent both bulk phase separation as well as a uniform wetting of the wall by a dense phase.
	
	\section{Localized deformation on a flat wall}\label{sec:multipole expansion}
	
	\subsection{Two dimensions}
\begin{figure}
		\centering
		\includegraphics[width=0.9\linewidth]{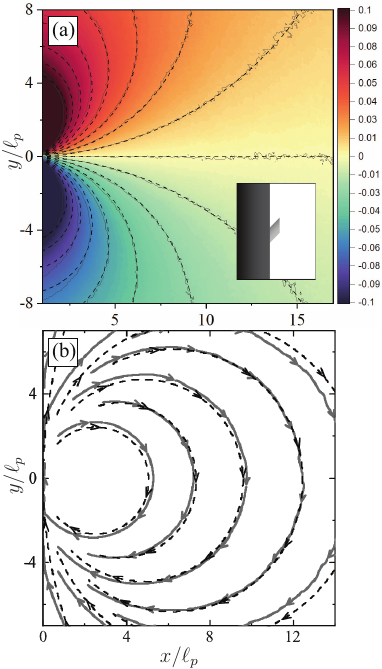}
		\caption{Density and current of RTPs near a flat wall in the presence of an isolated deformation.
			{\bf (a)} The color encodes the  density modulation $\phi(x,y)=\rho(x,y) - \langle \rho(x) \rangle$. The solid lines are contour lines plotted every $\delta \phi=1.25 \times 10^{-2}$ from the numerical data. They are compared with the corresponding theoretical predictions of Eq.~\eqref{eq:multipole expansion density}, shown by the dashed lines. ${\bf p}$ is measured numerically so that there is no fitting parameter.
			{\bf (b)} Streamlines of current measured in simulations (gray solid lines), compared to the theoretical prediction Eq.~\eqref{eq:dipolar current} (in dashed lines). For simulation details, see Appendix~\ref{app:simulation details}.}
		\label{fig:Single ratchet}
	\end{figure}

	In this section, we focus on the theoretical models of non-interacting Active Brownian Particles (ABPs) 
	and Run-and-Tumble Particles (RTPs).
	For simplicity, our calculations are carried out in two dimensions, and we present the results for higher dimensions in the next subsection. Each active particle follows the Langevin dynamics 
	\begin{align}\label{eq:Langevin ABPs RTPs 1}
		\frac{d{\bf r}_i}{dt} =& v\bfu(\theta_i) - \mu \nabla V\left({\bf r}_i\right) + \sqrt{2\mathcal{D}_t} \boldsymbol{\eta}_i\left(t\right)\ , \\
		\frac{d\theta_i}{dt} =& \sqrt{2\mathcal{D}_r}\xi_i(t)\ ,\label{eq:Langevin ABPs RTPs 2}
	\end{align}
	where ${\bf r}_i$ is the position of particle $i$, $v$ its self-propulsion speed, and $\bfu\left(\theta_i\right)=\left(\cos \theta_i ,\sin \theta_i \right)$ its orientation. The particle mobility is denoted as $\mu$ while  $\mathcal{D}_t$ and $\mathcal{D}_r$ are the translational and rotational noise amplitudes. Finally, $\boldsymbol{\eta}_i$ and $\xi_i$ are Gaussian white noises of unit variance and zero mean. In addition, the particle heading undergoes complete random reorientations, called tumbles, with rate $\alpha$.  ABPs and RTPs correspond to the limiting cases $\alpha=0$ and $\mathcal{D}_r=0$, respectively. The walls are modelled through the external potential $V({\bf r})$. 
	Our theoretical computations are carried out in a semi-infinite domain $x>0$ in the presence of a flat wall, perpendicular to the ${\bf \hat{x}}$ direction, assuming a bulk density $\rho_b$ at $x=+\infty$. An asymmetric obstacle of characteristic size $a$, representing a localized deformation of the wall, is located at $y=0$, as illustrated in the inset of Fig.~\ref{fig:Single ratchet}a. The obstacle is modelled as an additional potential $U(\bfr)$. 
	
	In this section we show that this deformation induces a steady-state density modulation whose far-field expression is given by:
	\begin{equation}\label{eq:multipole expansion density}
		\rho\left({\bf r}\right) \underset{x\gg a,\ell_p}{\simeq} \rho_b + \frac{\mu}{\pi \mathcal{D}_{\rm eff}} \frac{y p}{r^2} + \mathcal{O}({1}/{r^2})\ .
	\end{equation}
	Here, $r=\sqrt{x^2+y^2}$ is the distance from the deformation, $\mathcal{D}_{\rm eff}=\mathcal{D}_t+\frac{1}{2}v \ell_p$ is the effective diffusion coefficient, and $\ell_p=v/\left(\alpha+\mathcal{D}_r\right)$ is the particle's persistence length. The scale of the modulation is set by $p$, which measures the net force exerted by the obstacle on the active particles along the wall through
	\begin{equation}\label{eq:dipole moment}
		p = -\intop_0^\infty dx'\,\intop_{-\infty}^\infty dy'\,\rho\left({\bf r}'\right)\partial_y'U\ ,
	\end{equation}
	and is non-zero only for asymmetric obstacles. In the following, we refer to ${\bf p}\equiv p {\bf \hat y}$ as the \emph{force monopole} induced by the obstacle. The density modulation is accompanied by a current, which is diffusive in the far field ${\bf J}\simeq-\mathcal{D}_{\rm eff}\nabla \rho$, and is given by
	\begin{align}\label{eq:dipolar current}
		{\bf J}\left({\bf r}\right) \underset{x\gg a,\ell_p}{\simeq} \frac{\mu}{\pi}&\frac{2xy\hat{x} + \left(y^2-x^2\right)\hat{y}}{\left(x^2+y^2\right)^2}p + \mathcal{O}\left(1/r^3\right)\ .
	\end{align}
	Equation~\eqref{eq:dipolar current} predicts the flow created by a force monopole on the active fluid: It is the nonequilibrium diffusive counterpart of the Stokeslet flow in fluid dynamics, computed in the vicinity of a hard wall. Our results are verified and illustrated numerically in Fig.~\ref{fig:Single ratchet}. We now turn to their derivations, which are extended to homogeneous systems with pair-wise interactions in Appendix~\ref{app:multipole expansion interacting}.
	
	The probability density $\mathcal{P}({\bf r},\theta)$ to find an active particle located at ${\bf r}$ and oriented at an angle $\theta$ evolves according to the Master equation:
	\begin{align}\label{eq:FP}
		\partial_t \mathcal{P}({\bf r},\theta)= & -\nabla\cdot\left[v{\bf u}\mathcal{P}-\mu\nabla V\mathcal{P}-\mathcal{D}_t\nabla \mathcal{P}\right]+\mathcal{D}_r\partial_{\theta}^{2}\mathcal{P} \nonumber \\
		&-\alpha \mathcal{P}+\frac{\alpha}{2\pi}\intop d\theta'\,\mathcal{P}\left({\bf r},\theta'\right)\ .
	\end{align}
	For non-interacting particles, the average density field simply reads $\rho\left({\bf r}\right)= \intop d\theta\,\mathcal{P}\left({\bf r},\theta\right)$.  
	Integrating over $\theta$ leads to a conservation equation:
	\begin{equation}\label{eq:dynrho}
		\partial_t \rho({\bf r}) = - \nabla \cdot {\bf J}\;,
	\end{equation}
	where the current ${\bf J}$ is given by
	\begin{equation}\label{eq:currentJ}
		{\bf J}= v {\bf m} -\mu \rho \nabla V- \mathcal{D}_t \nabla \rho\;.
	\end{equation}
	It is the sum of a diffusive contribution due to translational noise, an advective current due to the external potential, and an active contribution proportional to ${\bf m} \equiv \int d\theta\, {\bf u}(\theta) \mathcal{P}({\bf r},\theta)$. Far away from the wall and the obstacle, the active dynamics is diffusive at large scales so that we expect $\bfJ \simeq -\mathcal{D}_{\rm eff} \nabla \rho$ in the steady state~\cite{cates_when_2013}. We can then introduce 
	\begin{equation}\label{eq:weirdJ}
		{\boldsymbol{\mathcal{J}}}\equiv {\bf J}+{\cal D}_{\rm eff} \nabla \rho\;,
	\end{equation}
	which measures the difference between $\bfJ$ and its bulk value to recast the conservation equation in the steady state, $\nabla \cdot {\bf J}=0$,  as
	\begin{align}\label{eq:Poisson's eq}
		\mathcal{D}_{\rm eff}\nabla^{2}\rho =\nabla \cdot \bfcJ\left({\bf r}\right)\;.
	\end{align}
	Equation~\eqref{eq:Poisson's eq} has the appealing feature of being a Poisson equation for the density field with a source term $\nabla\cdot \bfcJ({\bf r})$, which is expected to be non-vanishing only close to the wall and the deformation. This equation, however, has to be solved self-consistently since $\bfcJ$ depends on $\rho$ and $\bfm$. Furthermore, a second difficulty comes from the  non-trivial boundary condition imposed by the wall. Indeed, taking the limit of a hard wall, the component of the current transverse to the wall has to vanish, so that
	\begin{equation}\label{eq:BC}
		J_{x}\left(0,y\right)=\left(-\mathcal{D}_{\rm eff}\partial_{x}\rho
		+\mathcal{J}_x\right)\big|_{x=0}=0\ ,
	\end{equation}
	where $\mathcal{J}_x$ is the $x$-component of $\bfcJ$.
	This is neither a Dirichlet nor a Neumann boundary condition on $\rho$, since $\bfcJ$ is non-zero at the wall and depends on the density field.
	Nevertheless, since $\rho$ by itself is not prescribed on the boundary, we can still use the Neumann-Green's function of the Laplacian,
	\begin{equation}\label{eq:green}
		G_N({\bf r}_1,{\bf r}_2)=-\frac 1 {2\pi} [\ln(|\bfr_1-\bfr_2|)+\ln(|\bfr_1^\perp-\bfr_2|)]\;,
	\end{equation}
	to solve this boundary value problem. Here $\bfr^\perp\equiv (-x,y)$ is the image of $\bfr$ with respect to the wall. Using Green's second identity, one finds~\cite{jackson_classical_1999}
	\begin{align}
		\rho({\bf r}) =& -\frac{1}{\mathcal{D}_{\rm eff}}\intop_{0}^{\infty} dx'\intop_{-\infty}^{\infty}dy'\,  G_N(x,y;x',y')\nabla'\cdot \bfcJ' \nonumber \\
		&- \intop_{-\infty}^{\infty}dy'\,  G_N(x,y;0,y') \partial_x' \rho'\bigg|_{x'=0}+\rho_b \ ,\label{eq:general solution Poissons eq}
	\end{align}
	where $\partial_i'=\frac{\partial}{\partial r_i'}$ and $g'=g\left({\bf r}'\right)$ for any function $g({\bf r})$. Note that there are two important differences between the solution~\eqref{eq:general solution Poissons eq} and the density modulation that would be observed around an isolated obstacle in the bulk of an active fluid~\cite{baek_generic_2018}. First, the Green's functions differ between these two cases. Second, the surface integral in the second line of Eq.~\eqref{eq:general solution Poissons eq} would be absent in a bulk problem. Here, it ensures that no current flows through the wall. 
	
	Let us now analyze the behaviour of Eq.~\eqref{eq:general solution Poissons eq} in the far field, \emph{i.e.} when $|x-x'| \gg \ell_p,a$. We first split the divergence of $\bfcJ'$ as $\nabla' \cdot \bfcJ'=\partial_x' \mathcal{J}_x' + \partial_y' \mathcal{J}_y'$ and consider the contribution of $\partial_x' \mathcal{J}_x'$. Since $\bfcJ'$ is, to leading order, non-zero only close to the wall, the Green's function $G_N(x,y;x',y')$ can be expanded in $x'$ around $x'=0$:
	\begin{equation*}
		G_N(x,y;x',y')\simeq G_N(x,y;0,y')+\frac{{x'}^2}{2}\frac{\partial^2 G_N(x,y;0,y')}{\partial {x'}^2} \;,
	\end{equation*} 
	where have used ${\partial_x'}G_N(x,y;0,y')=0$ by symmetry. In the far field, $(x')^2(\partial_x')^2 G_N \ll G_N$ so that we neglect the second order derivative. The integral over $x'$ in Eq.~\eqref{eq:general solution Poissons eq} can then be carried out explicitly and, using Eq.~\eqref{eq:BC}, it directly balances with the surface integral, leading to
	\begin{equation}\label{eq:rhoint}
		\rho({\bf r})\!\!\! \underset{x\gg \ell_p,a}\simeq\!\!\!  \rho_b-\frac{1}{ \mathcal{D}_{\rm eff}}\intop_{0}^{\infty} dx'\intop_{-\infty}^{\infty}dy'\,  G_N(x,y;x',y')\partial_y' \mathcal{J}_y' \;.
	\end{equation}
	
	To evalute Eq.~\eqref{eq:rhoint}, we multiply~Eq.~\eqref{eq:FP} by $\bfu$ and integrate over $\theta$ to show that, in the steady state, $\frac{v}{\mu} \bfm= \nabla \cdot \sigma^a$, where
	\begin{equation}\label{eq:first moment}
		\sigma^a_{ij} =-\frac{\ell_p}\mu\left[\frac{v\delta_{ij}\rho}2 + v{Q}_{ij} -  (\mu\partial_j V+\mathcal{D}_t \partial_j) {m}_{i} \right]
	\end{equation}
	is known as the active pressure~\cite{Takatori2014,yang_aggregation_2014,Solon2015NatPhys,fily_mechanical_2017} and
	we have introduced $Q_{ij}(\bfr)\equiv \int d\theta\, (u_i u_j-\frac{\delta_{ij}}2)  \mathcal{P}(\bfr,\theta)$. From the definition of $\bfcJ$, one then has:
	\begin{equation}\label{eq:D'}
		\partial_y' \mathcal{J}_y'= \partial_y'[-\mu \rho \partial_y' V+\mu \partial_y' \sigma_{yy}'+\mu \partial_x' \sigma_{xy}'  +\frac{v \ell_p}{2} \partial_y' \rho]\;.
	\end{equation}
	To estimate the leading order contribution to the integral in Eq.~\eqref{eq:rhoint}, we use Eq.~\eqref{eq:D'} and integrate by parts. The three last terms in Eq.~\eqref{eq:D'} lead to two integrations by parts, hence involving the second order derivative of $G_N$. In the far field, they can, again, be neglected in comparison to the leading order term, which reads
	\begin{equation} \label{eq:rho_farfield}
		\rho({\bf r})\!\!\! \underset{x\gg \ell_p,a}\simeq\!\!\!  \rho_b-\frac{\mu}{ \mathcal{D}_{\rm eff}}\intop_{0}^{\infty} dx'\!\!\!\intop_{-\infty}^{\infty}dy'\,  \rho' \partial_y' U'\partial_y' G_N(x,y;x',y') \;.
	\end{equation}
	Here, we have used that $U$ is the only contribution to the potential that is not invariant by translation along $y$. Using the expression~\eqref{eq:green} for $G_N$ leads, to leading order in the far field, to Eq.~\eqref{eq:multipole expansion density}.
	
	Remarkably, while we embarked to solve the rather cumbersome problem posed by Eqs.~\eqref{eq:weirdJ} and \eqref{eq:Poisson's eq} with the boundary condition~\eqref{eq:BC}, the far-field solution~\eqref{eq:multipole expansion density} can be obtained by solving  a simpler problem:
	\begin{equation}\label{eq:poissonsimple}
		\mathcal{D}_{\rm eff} \nabla^2 \rho = -\mu \nabla \cdot [{\bf p} \delta(\bfr)]
	\end{equation}
	with ${\bf p}$ the force monopole exerted by the deformation and a standard Neumann boundary condition. 
	Thanks to this simplification, the problem of non-trivial boundaries can now be solved in higher dimensions and for more complex geometries with ease.
	
	\subsection{Higher dimensions}
	Using Eq.~\eqref{eq:poissonsimple}, or repeating the above calculation, in higher dimensions leads to 
	\begin{align}\label{eq:rho d dimensions}
		\rho\left({\bf r}\right)&\sim \rho_b + \frac{2\mu}{\mathcal{D}_{\rm eff} S_d} \frac{{\bf r}\cdot{\bf p}}{r^d} + \mathcal{O}\left(1/r^2\right) \\
		{\bf J}\left({\bf r}\right)&\sim \frac{2\mu}{S_d}\frac{d\left(\hat{r}\cdot{\bf p}\right)\hat{r}-{\bf p}}{r^{d}} + \mathcal{O}\left(1/r^2\right) \ ,\label{eq:J d dimensions}
	\end{align}
	where $S_d = (2\pi^\frac{d}{2})/\Gamma\left(\frac{d}{2}\right)$ and ${\bf p}$ is the force monopole exerted by the obstacle on the active particles along the wall:
	\begin{equation}
		{\bf p} = - \int d^{d}\bfr\, \rho(\bfr) \nabla_\parallel U(\bfr)\;.
	\end{equation}
	Here $\nabla_\parallel = \nabla - {\bf \hat{x}} \partial_x$ is the derivative operator acting parallel to the wall.
	Equations~\eqref{eq:rho d dimensions} and~\eqref{eq:J d dimensions} show that the density modulation and flows induced by a  localized deformation of a flat wall are solely controlled by the force monopole ${\bf p}$ exerted by the deformation on the active particles \textit{along} the wall,
	induced by asymmetry of the obstacles.

	\section{Disordered boundaries in dilute active systems}\label{sec: disordered boundaries}
	\label{sec: disordered boundaries: dilute systems}
	
	We now extend the results from an isolated deformation to the case of a disordered wall. The latter is modelled as a potential $V(x,\bfr_\parallel)$, where $\bfr_\parallel$ is a ($d-1$)-dimensional vector parallel to the wall. The potential is infinite for $x<0$ and is localized inside the interval $[0,x_w]$. In that region, $V(x,\bfr_\parallel)$ is drawn from a random, bounded distribution with a finite correlation length $a$. As we now show, the far-field modulation of the density field and the current generated by this disordered boundary are identical to those generated by force monopoles randomly placed along a flat wall and parallel to it. To do so, we first compute analytically the current and density modulations created by such random force monopoles and later compare them with microscopic numerical simulations.
	
	\subsection{Long-range density correlations}
	
	Consider a continuous, quenched, Gaussian random variable, ${\bf f}({\bf r}_\parallel)$, describing the force-monopole density along the wall, whose disorder-average satisfies:
	\begin{align}\label{eq:stat}
		&\overline{f_i\left({\bf r}_\parallel\right)} = 0\ ,\nonumber \\
		&\overline{f_i\left({\bf r}_\parallel\right)f_j\left({\bf r}_\parallel\!'\right)} = 2 p^2\delta_{ij}^{\parallel} \sigma^2\delta^{(d-1)}\left({\bf r}_\parallel-{\bf r}_\parallel\!'\right)\ ,
	\end{align}
	with $p$ setting the scale of the force, $\sigma^2\simeq a^{1-d}$ an inverse area related to the microscopic correlation length of $V$, $\delta^\parallel_{ij}=1$ if $i=j \neq x$, and $\delta^\parallel_{ij}=0$ otherwise. To determine the density modulations, we rely on Eq.~\eqref{eq:poissonsimple} and solve
	\begin{equation} 
		{\cal D}_{\rm eff} \nabla^2 \rho = - \mu \nabla \cdot [{\bf f} ({\vec r}_\parallel) \delta(x)]\,,
	\end{equation}
	with a Neumann boundary condition. In the far field, this leads to: 
	\begin{equation}\label{eq:density with dipole density}
		\rho\left(x,{\bf r}_\parallel\right) \approx \rho_b + \frac{2\mu}{\mathcal{D}_{\rm eff} S_d} \intop d^{d-1}r_\parallel\!' \, \frac{\left({\bf r}_\parallel - {\bf r}_\parallel\!' \right)\cdot {\bf f}\left( {\bf r}_\parallel\!' \right)}{\left[x^2 + \left|{\bf r}_\parallel - {\bf r}_\parallel\!'\right|^2\right]^{\frac{d}{2}}} \ .
	\end{equation}
	
	We first note that, on average, $\overline{\rho\left({\bf r}\right)}\approx\rho_b$ in the far field: a disordered wall thus does not generate a systematic density modulation in the far field. However, a non-trivial structure is revealed by computing the disorder-averaged two-point connected correlation function:
	\begin{equation}\label{eq:rhorho}
		\overline{\rho(x,{\bf r}_\parallel) \rho(x',{\bf r}_\parallel\!')}_c = \frac{1}{S_d}\!\left(\frac{2\mu p \sigma}{\mathcal{D}_{\rm eff}}\right)^2\!\!\frac{\left(x+x'\right)}{\left[\left(x+x'\right)^2 + \left|\Delta{\bf r}_\parallel\right|^2\right]^\frac{d}{2}},
	\end{equation}
	where $\Delta{\bf r}_\parallel = {\bf r}_\parallel - {\bf r}_\parallel\!'$. This equation predicts large-scale density modulations which decay in amplitude---but increase in range---as one moves away from the wall. To see this, consider the case in which $x=x'$. For $\Delta{\bf r}_\parallel=0$, the two-point function decays as $\overline{\rho(x,{\bf r}_\parallel) \rho(x,{\bf r}_\parallel\!)}_c \sim 1/x^{d-1}$, showing that the disorder-induced density fluctuations are stronger close to the wall. The transverse correlations of these fluctuations, however, only decay when $|\Delta{\bf r}_\parallel|\gg 2 x$: their correlation length thus \textit{increases} with the distance from the wall. 
	
\begin{figure}
		\includegraphics[width=0.9\linewidth]{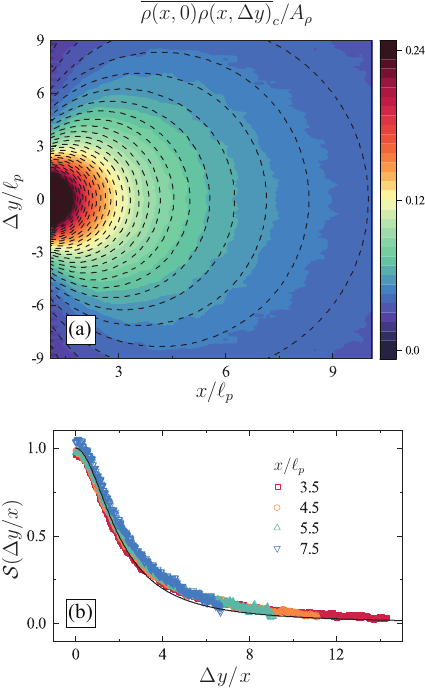}
		\caption{
			Disorder-averaged two-point density correlation function of non-interacting RTPs in two-dimensions in the presence of a disordered wall at $x=0$. {\bf (a)} The two-point correlation function as $x$ and $\Delta y$ are varied, calculated from simulations, is shown by the color map. 
			The value of $A_\rho \equiv \ell_p S_d^{-1} (2 \mu p \sigma / {\cal D}_\mathrm{eff})^2$ is obtained from
			a fit of the data to Eq.~\eqref{eq:rhorho}. The latter includes a constant offset due to finite-size corrections, which is calculated exactly in Appendix~\ref{app:periodic walls}. 
			The theoretical prediction of Eq.~\eqref{eq:rhorho} is then used to produce dashed contour lines that match the levels of the color bar.
			Both theory and simulations are normalized by $A_\rho$.
			{\bf (b)} A verification of the scaling form~\eqref{eq:C} for the density-density correlation function. The data shown in panel (a) for four different distances $x$ from the wall are collapsed onto a single curve, as predicted.  See Appendix~\ref{app:simulation details} for numerical details.
		}
		\label{fig:nonint_scaling}
	\end{figure}

	These results are qualitatively illustrated and quantitatively checked in Fig.~\ref{fig:nonint_scaling} using microscopic simulations which demonstrate the relevance of the model~\eqref{eq:stat} for disordered boundaries. First, we measure numerically $\overline{\rho(x,{\bf r}_\parallel) \rho(x,{\bf r}_\parallel\!')}_c$ which we fit against the right-hand side of Eq.~\eqref{eq:rhorho} to extract the value of $\sigma$. The numerical data, normalized by the prefactor $4\mu^2 p^2 \sigma^2/(S_d D_{\rm eff})$, are then shown to match the contour lines predicted by Eq.~\eqref{eq:rhorho}. A more quantitative comparison can be obtained by noticing that the correlation function can be rescaled as: 
	\begin{align}
		\frac{\overline{\rho(x, y) \rho(x,y+\Delta y)}_c}{\overline{\rho(x, y) \rho(x,y)}_c}
		&= \frac{1}{1 + \left( \frac{\Delta y}{2x} \right)^2 }\equiv \mathcal{S} \left( \frac{\Delta y}{x} \right)\;, \label{eq:C}
	\end{align}
	leading to a scaling form.  Figure~\ref{fig:nonint_scaling}(b) shows the quantitative agreement between the numerical data and the prediction of Eq.~\eqref{eq:C}. 
	
	\subsection{Current cascade}
\begin{figure}
		\centering
		\includegraphics[width=0.9\linewidth]{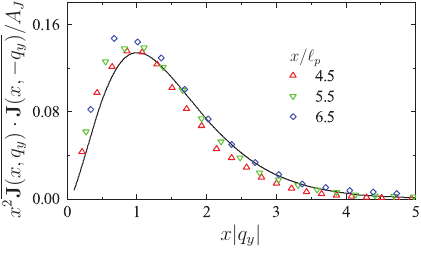}
		\caption{Fourier transform along the $\hat {\bf y}$ direction of the current-current correlation function measured at a distance $x$ from the wall and averaged over disorder.  The  data are measured for three values of $x$ and normalized by a factor $A_J \equiv 2 d (2 \pi)^{d-1} (\mu \sigma p)^2$. As predicted by our theory, the data can be collapsed onto a single curve, corresponding to Eq.~\eqref{eq:current correlations}, by properly scaling the abscissa and the ordinates.  See Appendix~\ref{app:simulation details} for numerical details.
		}
		\label{fig:current_scaling}
	\end{figure}

	Another interesting way to interpret these results is to consider the impact of the disordered boundary on the particle current. On a microscopic scale close to the wall, the random forcing induced by the disorder stirs the active medium. The conservation law for the density field then turns this microscopic stirring into large-scale eddies in the bulk of the system.
	This cascade structure can be quantified by analysing the statistics of the steady-state currents. 
	In the bulk of the system, the large-scale current can be estimated as~\cite{cates_when_2013}:
	\begin{equation}\label{eq:current d dimensions disorder}
		{\bf J}\left(x,{\bf r}_\parallel\right) \approx -\mathcal{D}_{\rm eff} \nabla \rho\left(x,{\bf r}_\parallel\right)\ .
	\end{equation}
	Using Eqs.~\eqref{eq:density with dipole density} and~\eqref{eq:current d dimensions disorder}, and
	performing a Fourier transform with respect to ${\bf r}_\parallel$, leads to:
	\begin{eqnarray}
		J_x\left(x,{\bf q_\parallel}\right) &=& -i\mu {\bf q_\parallel}\cdot{\bf f}_{{\bf q_\parallel}}\, e^{-\left|{\bf q_\parallel}\right|x}\\
		J_k \left(x,{\bf q_\parallel}\right) &=& \text{sign}(q_{_\parallel,k})\mu {\bf q_\parallel}\cdot{\bf f}_{{\bf q_\parallel}}\, e^{-\left|{\bf q_\parallel}\right|x}
	\end{eqnarray}
	where $k$ describes one of the $d-1$ dimensions parallel to the wall and ${\bf f}_{{\bf q_\parallel}} \equiv \int d^{d-1}r_\parallel {\bf f}({\bf r}_\parallel)e^{-i {\bf q_\parallel}\cdot{\bf r}_\parallel}$. Taking a disorder average and using Eq.~\eqref{eq:stat} then leads to
	\begin{align}\label{eq:current correlations}
		\overline{{\bf J}\left(x,{\bf q_\parallel}\right) \cdot {\bf J}^*\left(x,{\bf q_\parallel}'\right)} =& \,2d \left(\mu \sigma p\right)^2 \left|{\bf q_\parallel}\right|^2 e^{-2\left|{\bf q_\parallel}\right| x} \times \nonumber \\
		& \times \left(2\pi\right)^{d-1}\delta^{(d-1)}\left({\bf q_\parallel}+{\bf q_\parallel}'\right)\ .
	\end{align}
	This result shows that, for a given value of $x$, the current-current correlations first increase for small $|{\bf q_\parallel}|$ before they are exponentially suppressed by the term $\exp(-2 |{\bf q_\parallel}| x)$. The larger the value of $x$, the smaller the values of $|{\bf q_\parallel}|$ for which the peak of the correlation function is observed, revealing eddies on larger and larger scales as $x$ increases. This explains the large-scale structures exhibited by the current in Fig~\ref{fig:rho_and_jy}. Our predictions~\eqref{eq:current correlations} are verified quantitatively in Fig.~\ref{fig:current_scaling}, using a scaling form similar to that of Eq.~\eqref{eq:C}.
	
	\subsection{Other geometries}
	
	The methodology presented above can be extended to other boundary shapes. For instance, a corrugated border that repeats periodically along the $\hat {\bf y}$ direction is studied in Appendix~\ref{app:periodic walls}. Our analytical results show the large-scale density-density correlations to be exponentially suppressed at a distance corresponding to the periodicity of the potential. This explains why localized currents had been reported in the presence of periodic asymmetric walls~\cite{nikola_active_2016}, instead of the cascade structure revealed in the previous section.
	
\begin{figure}
		\includegraphics[width=0.9\linewidth]{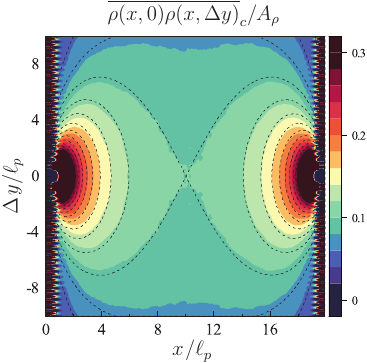}
		\caption{
			Disorder-averaged two-point density correlation function of non-interacting RTPs measured in the presence of two disordered walls at $x=0$ and $x=20\ell_p$. Periodic boundary conditions are imposed along the ${\bf \hat{y}}$ direction. The correlation function is normalized by a factor $A_\rho \equiv \ell_p S_d^{-1} (2 \mu p \sigma / {\cal D}_\mathrm{eff})^2$. Simulation results are shown as a color map and compared to the analytic predictions of Eq.~\eqref{eq:rhorho two walls} (dashed contour lines). See Appendix~\ref{app:simulation details} for numerical details.}
		\label{fig:nonint_two_wall}
	\end{figure}

	Another important case pertains to multiple interfering boundaries. For example, Figure~\ref{fig:nonint_two_wall} shows the disorder-averaged correlation functions at $x'=x$ for non-interacting RTPs between two disordered walls, with a periodic boundary condition in the ${\bf \hat y}$ direction. The analytic expression for the correlation function is calculated and given in Appendix~\ref{app:two walls}. In the bulk of the system, the interplay between the two walls leads to a decrease of the transverse correlations and to their suppression in the vicinity of $x=L_x/2$. This highlights how boundaries can control the bulk behaviours of active systems as well as the importance of properly including them in the theoretical description of active matter.

	\section{Disordered boundaries in interacting active systems}\label{sec: disordered boundaries: interacting systems}
	
	To study the influence of disordered boundaries on interacting active-matter systems, we rely on a linear field theory that builds on the force-monopole picture presented above. Our results are then validated using a self-consistency argument and by the explicit comparison with microscopic numerical simulations. 
	
	\subsection{Linear field theory}
	To proceed, we consider a system of active particles at an average density $\rho_b$ in $d$ space dimensions and consider the density-fluctuation field $\phi({\bf r}) \equiv \rho({\vec r}) - \rho_b$. The particles are in contact with a $d-1$ dimensional wall with a random potential along it. Since the number of particles is conserved, $\phi({\bf r})$ undergoes model-B type dynamics 
	\begin{align} \label{eq:field}
		\partial_t \phi({\bf r}, t) &= - \nabla \cdot {\bf J} 	({\bf r}, t)\;, \\
		{\bf J}( {\bf r}, t) &= - \nabla g [\phi] + {\bf f} ({\bf r}) +\sqrt{2D} \boldsymbol{\eta} ({\bf r}, t)~.\label{eq:J}
	\end{align}
	Here, ${\bf J} 	({\bf r}, t)$ is a current and $g[\phi]$ plays the role of a chemical potential. We first consider a linear theory in  which
	\begin{equation} \label{eq:field_linear}
		g [\phi ({\bf r}, t)] = u \phi ({\bf r}, t) - K \nabla^2 \phi ({\bf r}, t)\;,
	\end{equation}
	where $\boldsymbol{\eta} ({\bf r}, t)$ is a unit Gaussian white-noise field satisfying
	\begin{equation}\label{eq:eta}
		\langle \eta_i({\bf r},t)\eta_j({\bf r'},t') \rangle = \delta_{ij}\delta^d({\bf r}-{\bf r'})\delta(t-t') \; ,
	\end{equation}
	the mobility has been set to be one, and $K>0$ for stability. As argued in the previous section, on a coarse-grained scale, the quenched random potential of the boundary amounts to a random force field {\it along} the wall. We account for it through a quenched random force-density field ${\bf f}({\bf r})$ that is parallel to the wall and satisfies
	\begin{align}
		{f}_x (x,{\bf r}_\parallel) =&\, 0\;, \label{eq:fstat} \\
		{ \overline{{f}_i (x, {\bf r}_\parallel)} =}&{ \, 0\;,} \\
		\nonumber \overline{ {f}_i (x, {\bf r}_\parallel) {f}_j (x', {\bf r'}_\parallel ) } =&\, 2 s^2 \delta_{ij} \delta(x) \delta(x') \delta^{(d-1)} ({\bf r}_\parallel - {\bf r'}_\parallel)\;,
	\end{align}
	where $i$ and $j$ label directions parallel to the wall. Note that, in contrast to Eq.~\eqref{eq:stat}, we have included the factor $\delta(x)$ in the definition of ${\bf f}(\bfr)$. Finally, the strength $s$ of the random force is allowed to depend on $\rho_b$ but is, to leading order, independent of $\phi$.
	\begin{figure}
		\includegraphics[width=0.9\linewidth]{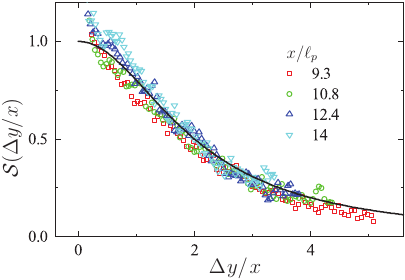}
		\caption{Scaled density-density correlation function defined in Eq.~\eqref{eq:C} for interacting RTPs. The simulation results, shown in symbols, are obtained by varying $\Delta y$ at fixed $x$. The solid line corresponds to the theoretical prediction of Eq.~\eqref{eq:C}.  See Appendix~\ref{app:simulation details} for numerical details.
		}
		\label{fig:int_cor}
	\end{figure}
	
	As detailed in Appendix~\ref{app:structure factor}, the structure factor $S({\bf q}, {\bf q}') \equiv \overline{ \langle \phi({\bf q} ) \phi({{\bf q}}') }\rangle$ can be directly evaluated, leading to:
	\begin{align} 
		S({\bf q}, {\bf q}') =&  \frac{ 2 s^2  (2 \pi)^{d-1}  |{\bf q}_\parallel|^2 \delta^{(d-1)} ( {\bf q}_\parallel + {\bf q}_\parallel')  }{ q^2 q'^2 (u + Kq^2) (u + K q'^2)} \nonumber \\
		&+ \frac{2 D (2 \pi)^{d-1} \delta^d ({\bf q} + {\bf q}') }{ (u + Kq^2)^2 } \label{eq:structure_factor}
	\end{align}
	where the brackets denote a steady-state average. Interestingly, the long-wavelength behavior is controlled by the random forcing term so that the small $q$ behavior is given by
	\begin{equation} \label{eq:structure_asymptotic}
		S({\bf q}, {\bf q}') \sim  (2 \pi)^{d-1}  \frac{2 s^2 |{\bf q}_\parallel|^2}{u^2(q q')^{2}}\delta^d ( {\bf q}_\parallel + {\bf q}_\parallel') \;.
	\end{equation}
	In particular, in the limit ${q}^2, {q}'^2 \ll u/K$, the correlation function $\overline{ \langle \phi({\bf r} ) \phi({{\bf r}}') \rangle}$---obtained by performing an inverse Fourier transform on Eq.~\eqref{eq:structure_factor}---agrees with Eq.~\eqref{eq:rhorho}. This allows us to identify $s/u = 2 \mu p \sigma / {\cal D}_\mathrm{eff}$ as the strength of the random forcing in the dilute regime. 
	
	\subsection{Self-consistency of the linear field-theory.}
	
	We now check the self-consistency of our linear theory against the addition of non-linear terms in $g[\phi]$. To do this, we consider
	\begin{equation} \label{eq:field_nonlinear}
		g[\phi({\bf r}, t)] = u \phi({\bf r}, t) - K \nabla^2 \phi({\bf r}, t) + g \phi^n({\bf r}, t),
	\end{equation}
	with $n \geq 2$, and examine the scaling of the coefficient of $g$ under the rescaling
	\begin{eqnarray} \label{eq:rescaling}
		{\bf r} \to b {\bf r}, \quad t \to b^z t, \quad \phi \to b^\chi \phi \;.
	\end{eqnarray}
	The dynamic exponent $z=2$ is diffusive~\footnote{This can be confirmed, for instance, using the two-point two-time correlation function of the density field. In the large $b$ limit, it admits a non-trivial scaling form only if $z=2$.}.  
	At the fixed point of the linear theory, Eq.~\eqref{eq:structure_asymptotic} has to be preserved under rescaling. The coupling $(s^2/u^2)$ in Eq.~\eqref{eq:structure_asymptotic} renormalizes as
	\begin{equation}
		\left( \frac{s^2}{u^2} \right)' = b^{-2 \chi -d + 1} \left( \frac{s^2}{u^2} \right) \;, 
	\end{equation}
	which sets
	\begin{equation}
		\chi = \frac{1-d}{2}.
	\end{equation}
	The non-linearity is thus rescaled as $g \to b^{(n-1)(1-d)/2 }g$. For $d>1$, the term $g \phi^n$ is irrelevant. Note that, consistent with the result of the previous subsection, the term $K \nabla^2 \phi$ is also irrelevant, as would any higher order gradient terms like  $(\nabla \phi)^2$. All in all, the linear theory is thus self-consistent for $d>1$. We now turn to the numerical verification of Eq.~\eqref{eq:structure_asymptotic} using microscopic simulations of interacting active particles.
	
	\subsection{Numerical results}
	
	We performed numerical simulations of the microscopic active lattice gas described in Appendix~\ref{app:simulation details} in the presence of partial exclusion. The scaling form of the correlation function~\eqref{eq:C} is verified numerically in Fig.~\ref{fig:int_cor}. The boundary-induced long-ranged correlations revealed in dilute active systems are thus robust to the addition of interactions, hence validating our linear field theory.

	The latter describes active systems as long as the density field remains the sole hydrodynamic field. As such, the large-noise disordered phases encountered in the presence of aligning interactions, whether polar or nematic, will exhibit a similar behavior. In particular, this means that the bulk large-scale behavior of scalar active matter in the presence of disordered boundaries is controlled by the boundary and \textit{not} by particle interactions. 
	
	Our results suggest that the studies of bulk phase transitions of scalar active systems are likely to yield different results depending on the type of boundaries. Unlike in equilibrium systems, the generalization of results obtained in the presence of periodic boundaries should thus be questioned. To this end, in the next section we study the fate of motility-induced phase separation in the presence of disordered walls.
	
	\section{The effect of disordered boundaries on MIPS}\label{sec:Imry-Ma}
	
	In equilibrium, it is known that liquid-gas phase separation is completely unaffected by the presence of disorder on the boundaries of the system~\cite{lebowitz1999statistical}. Their contribution to the free energy is indeed sub-extensive so that it has no influence on the system's bulk behavior. In this section we show that, for scalar active systems, the situation is dramatically different: the long-ranged density modulations induced by the disordered boundaries lead to the suppression of bulk phase separation in any dimensions $d<d_c$ with $d_c=3$.

	To show this, we rely on our linear field theory, Eqs.~\eqref{eq:field}-\eqref{eq:field_linear} and use a Helmholtz-Hodge decomposition of the random forcing:
	\begin{equation}
		{\bf f}({\bf r}) = - \nabla U({\bf r}) + {\bf j} ({\bf r})  \;.
	\end{equation}
	We identify $U(\bfr)$ as an effective potential while ${\bf j} ({\bf r})$ captures the divergence-free part of the force field. The dynamics of Eq.~\eqref{eq:field} then implies that the statistics of the density field are insensitive to ${\bf j} ({\bf r})$. Scalar active systems with disordered boundaries thus share the bulk behaviour of a passive equilibrium problem with an effective potential $U(\bfr)$ that we now characterize. 
	
	By definition, the effective potential satisfies $\nabla^2 U({\bf r}) = - \nabla \cdot {\bf f} ({\bf r})$. Using Eq.~\eqref{eq:fstat}, it is then straightforward to show that the effective potential obeys
	\begin{align} \label{eq:U}
		\overline{U({\bf r})} &= 0\;, \\
		\overline{U({\bf r}) U({\bf r}') } &= \frac{s^2}{S_d} \frac{(x + x')}{\left[ (x + x')^2 + | \Delta {\bf r}_\parallel|^2 \right]^{d/2}}\;.
	\end{align}
	
	\begin{figure}
		\includegraphics[width=0.8\linewidth]{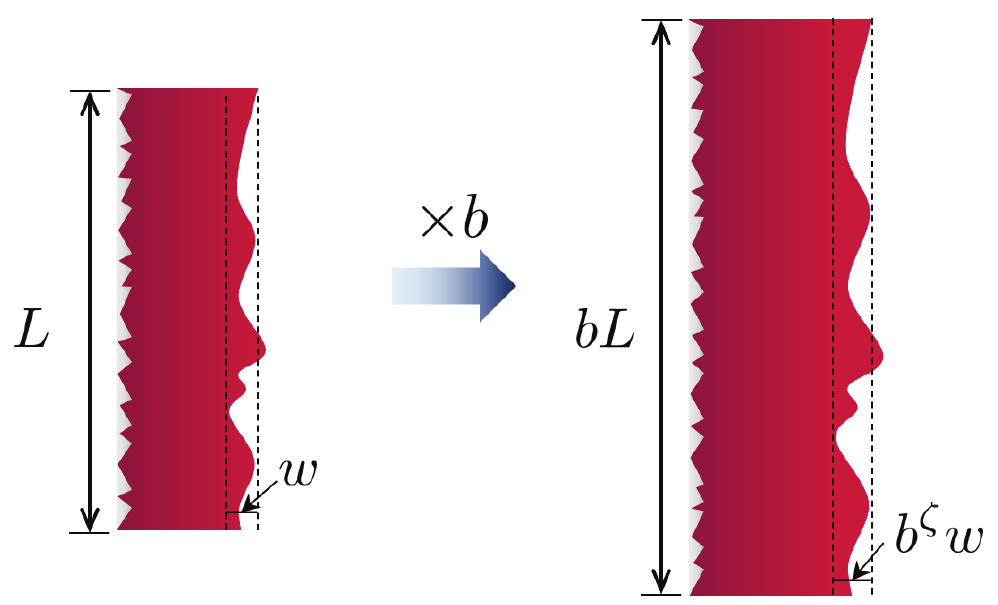}
		\caption{An illustration of the scaling procedure used to construct the Imry-Ma argument. As the system size is increased by a factor of $b$, the width of the interface between the phases is multiplied by $b^\zeta$. The interface is well defined in the large system-size limit when $\zeta<1$.}
		\label{fig:scaling}
	\end{figure}
	
\begin{figure*}
		\includegraphics[width=0.66\linewidth]{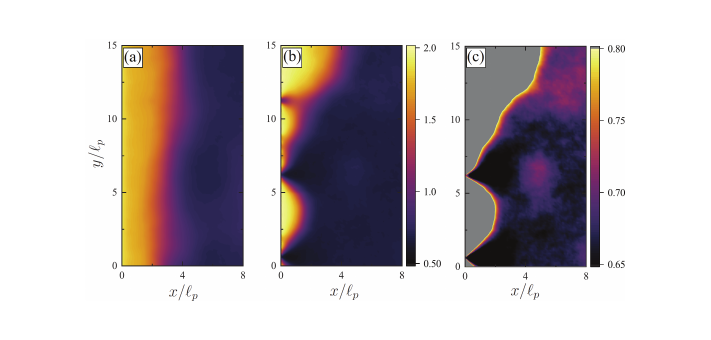}
		\caption{Time-averaged density of interacting RTPs. {\bf (a)} Density field in the presence of a hard flat wall at $x=0$, in the absence of  disorder. {\bf (b)} The density field in the presence of disorder along the wall. The uniform wetting layer shown in panel (a) is broken into random patches of varying size that prevent macroscopic phase separation. (c) The same as (b) with a smaller density range that reveals the long-ranged density modulations in the bulk of the system.}
		\label{fig:Imry-Ma}
	\end{figure*}

	With this in mind, we construct an Imry-Ma argument~\cite{imry_random-field_1975,aharony_lowering_1976,berker_ordering_1984} to determine when a phase-separated profile is stable against boundary disorder. It is well known that active particles tend to wet hard boundaries so that the liquid phase is usually  localized in their vicinity (see Fig.~\ref{fig:Imry-Ma}(a)). We thus study the fate of a macroscopic, fully wetting layer of the liquid phase when increasing the system size. Alternatively, we discuss the case of a macroscopic liquid droplet in the bulk of the system in Appendix~\ref{app:Imry-Ma}, which leads to identical conclusions.  
	
	To examine the stability of the wetting configurations, we study the roughness of the interface separating the dense and dilute phases~\cite{kardar1987domain}. 
	Its location is described by a height function $h({\bf r}_\parallel)$, with ${\bf r}_\parallel$ being the coordinate along the wall. 
	Upon rescaling the system size ${\bf r} \to b {\bf r}$, the interface width scales as $w \to b^\zeta w$. For a phase-separated configuration to be macroscopically stable, the roughness exponent must satisfy $\zeta < 1$. Otherwise, the existence of a well-defined interface is not self-consistent. 
	
	To compute $\zeta$, we consider an interface fluctuating around a mean height $h_0$. The elastic contribution of the interface to the free energy is given by
	\begin{equation} \label{eq:e_gamma_roughening}
		E_\gamma = \int_{L^{d-1}} d^{d-1} {\bf r}_\parallel ~ \left[ \frac{\gamma}{2} \left( \nabla h ({\bf r}_\parallel) \right)^2 \right],
	\end{equation}
	while the change due to the effective potential reads:
	\begin{equation} \label{eq:e_u_roughening}
		E_U = \int_{L^{d-1}} d^{d-1} {\bf r}_\parallel \int_0^{\delta h({\bf r}_\parallel)} d h' ~ \left[ \rho_0 U({\bf r}_\parallel, h_0 + h') \right],
	\end{equation}
	where $\gamma$ is the stiffness of the interface and $\delta h({\bf r}_\parallel) \equiv h({\bf r}_\parallel) - h_0$. To proceed, we compare the scalings of $E_\gamma$ and $E_U$ upon multiplying the system size by a factor of $b$. By definition, the latter implies $h_0 \to b h_0$ and $\delta h \to b^\zeta \delta h$. Inspection of Eq.~\eqref{eq:e_gamma_roughening} shows that $E_\gamma$ is rescaled as $E_\gamma \to b^{2 \zeta + d - 3} E_\gamma$. The scale of $E_U$ can be estimated from $|E_U| \equiv \sqrt{\overline{E^2_U}}$, which leads to $E_U \to b^{(d-1+2 \zeta)/2} E_U$.
	
	In a phase-separated system, where the interface is well-defined, its fluctuations are set by the balance between $E_\gamma$ and $E_U$. This requires matching their scaling exponents, which leads to
	\begin{equation}
		\zeta = \frac{5-d}{2}~.
	\end{equation}
	Importantly, phase separation with a smooth interface requires $\zeta<1$, which is only possible for $d>3$.  For dimensions $d < d_c$ with $d_c = 3$, the  width of the interface would diverge faster than the size $h_0$ of the domain: phase separation is no longer possible.  MIPS is thus  unstable against boundary disorder for dimensions $d< d_c$ with $d_c = 3$.
	
	Our predictions above are demonstrated numerically in Fig.~\ref{fig:Imry-Ma} using interacting RTPs on lattice in $d=2$. In Figs.~\ref{fig:Imry-Ma}a and~\ref{fig:Imry-Ma}b, we compare the steady-state densities of RTPs with and without disorder along the wall. (See Appendix~\ref{app:simulation details} for details.) In the absence of disorder, a stable phase separation is observed in the form a macroscopic, fully-wetting layer. In contrast, in the presence of disorder along the wall, a broken interface is observed, consistent with our Imry-Ma argument. Closer inspection of the bulk, shown in Fig.~\ref{fig:Imry-Ma}c, reveals large-scale correlations reminiscent of the non-interacting case. Indeed, as predicted, the density field in the bulk exhibits long-ranged correlations consistent with Eqs.~\eqref{eq:rhorho} and \eqref{eq:structure_asymptotic}. This is shown in Fig.~\ref{fig:int_cor}.

	Finally, to illustrate dynamically how wall disorder suppresses phase separation in the bulk of the system, we report in SM Movie 1 the following numerical experiment. A system is simulated in the presence of flat walls in the absence of wall disorder, leading to a macroscopic phase separation. To complement the above discussion, we choose parameters such that the macroscopic liquid droplet is deep in the bulk of the system. Then, the flat walls are replaced by disordered ones and the system is let to relax. The bulk droplet evaporates and is randomly redistributed across the system, consistent with the Imry-Ma argument of Appendix~\ref{app:Imry-Ma}.
	
	\section{Conclusions}
	
	In this work, we have shown that disordered boundaries exert a surprising influence on the bulk of active systems, leading to long-ranged correlations, current cascades, and the destruction of bulk phase separation. Our results are valid for scalar active matter and are robust to interactions between the particles as long as density remains the sole hydrodynamic field. This strongly differs from equilibrium systems in which the influence of boundaries can generically be discarded (for an interesting exception, see~\cite{feldman2002destruction}).  Our results were derived for RTPs and ABPs, but they can be straightforwardly extended to other classes of active particles like active Ornstein-Uhlenbeck particles~\cite{szamel2014self,martin2021statistical}.
	
	Experimentally, the sensitivity of active matter to boundaries has attracted a lot of attention in the past~\cite{kudrolli2008swarming,galajda_wall_2007}. In response, many boundary designs have been suggested to suppress their impact on the system~\cite{Deseigne2010PRL}. Our work shows that boundary effects are not restricted to finite-size systems and would persist in the thermodynamic limit. By offering a quantitative way to account for the influence of boundaries, we instead raise the question as to how boundaries can be used to control the bulk properties of active systems. Answering this challenging question will require adapting the methodology developed in this article to more general boundary shapes. 
	
	\section{Acknowledgments}
	We thanks Ari Turner for useful discussions. YBD, SR, YK and MK were supported by an NSF5-BSF grant (DMR-170828008). YBD, SR, YK acknowledge support from an ISF grant. JT acknowledges support from ANR grant THEMA. All authors benefited from participation in the 2020 KITP program on Active Matter supported by the grant NSF PHY-1748958.

	\appendix
	
	\section{Numerical simulations}\label{app:simulation details}
	\setcounter{equation}{0}
	
	All our numerics on active systems correspond to RTPs in two dimensions. Our theoretical predictions were successfully tested against both off-lattice and on-lattice simulations. In this article, we solely report the latter for which larger sizes and times can be reached. 
	
	\if{
		\noindent {\it Off-lattice simulations -- a localized deformation on a wall:} We consider $N$ particles on a system of dimensions $L_x \times L_y$ with periodic boundary conditions along the $\hat{\bf y}$ direction and hard walls at $x=0$ and $x=L_x$. The deformation on the boundary is modelled using a hard rod potential, see Fig.~\ref{fig:Single ratchet}a, of length $\ell_r$ whose angle with the wall is $\varphi$.
		
		We simulate RTPs moving according to Eqs.~\eqref{eq:Langevin ABPs RTPs 1} and \eqref{eq:Langevin ABPs RTPs 2} with ${\cal D}_t={\cal D}_r=0$. To this end, each particle is assigned an orientation $\bfu (\theta) = (\cos \theta, \sin \theta)$ and tumbles to a new  random orientation sampled uniformly in $\theta\in[0,2\pi)$ with a rate $\alpha$. To this end, the times between two successive tumbles are sampled from an exponential law of mean $\alpha^{-1}$. Between any two consecutive tumbles, the position of each particle is updated as follows. In free space, the particle position evolves as
		\begin{align}\label{eq:dynoffL}
			\bfr_i (t_{n+1}) = \bfr_i (t_{n}) + v\bfu \left(\theta_i(t_n)\right) \Delta t\;,
		\end{align}
		where $\bfr_i (t_{n})$ is the position of the $i$-th particle at time $t_n$, $\Delta t \equiv t_{n+1} - t_n$ is the time step. In case the particle collides with the wall or the rod, its equation of motion is integrated analytically as follows: Using Eq.~\eqref{eq:dynoffL} until the particles collide with the wall, and using a self-propulsion velocity $v P \cdot \bfu$ after the collision, where $P={\bf e}_\parallel {\bf e_\parallel}$ is the projector parallel to the wall. When particles reach the corner between the wall and the rod, they stop their motion until a tumble occurs. The force exerted by the particles on the rod is calculated as $|v {\bf u}\cdot {\bf e_\perp}/\mu|$, where ${\bf e_\perp}$ is the unit vector normal to the wall. This measurement is used for plotting the theoretical equal-density lines of Fig.~\ref{fig:Single ratchet}a.
		
		To measure the density and current at each point in space, we use a grid and time-average the number of particles and the particle current in each grid cell. Each grid cell has a linear size $\Delta$. The current through the grid cell ${\bf i}\equiv (i,j)$ is measured as $\bfJ_{{\bf i}}(t_n) = \sum_{k=1}^N \intop_{x_{{\bf i}}}^{x_{{\bf i}}+\Delta } dx  \intop_{y_{{\bf i}}}^{y_{{\bf i}}+\Delta } dy\, \delta(x-x_k(t_n)) \delta(y-y_k(t_n))[\bfr_k(t_n) - \bfr_k(t_{n-1})]/\Delta t$, where $x_{\bf i}=i \Delta$ and $y_{\bf i}=j \Delta$. The current streamlines displayed in Fig.~\ref{fig:Single ratchet}b are obtained using the streamslice Matlab tool. The parameters used to produce Fig.~\ref{fig:Single ratchet} are: $N = 1.5 \times 10^3$, $v = 5 \times 10^{-4}$, $\alpha = 10^{-2}$, $\Delta t = 1$, the measurement time is $t_0=10^9$, $L_x = 20 \ell_p$, $L_y = 60 \ell_p$, $\varphi=\frac{\pi}8$, $\ell_r=\ell_p$. Here, as before, $\ell_p = {v}/{\alpha}$. The grid cell size is $\Delta=0.1 \ell_p$.
	}\fi
	
	\smallskip
	
	We consider $N$ RTPs with and without interactions on a two-dimensional lattice of size $L_x \times L_y$. The system is periodic along the $\hat{\bf y}$ direction and confined by hard walls at $x<0$ and $x\geq L_x$. 
	
	\textit{Disordered wall:}
	The quenched disordered potential is modelled by placing wedge-shaped asymmetric obstacles along the wall, at every $\delta y=\ell_w$, whose orientations are chosen randomly (See Fig~\ref{fig:schematic} for a qualitative illustration). The obstacles have a finite extent $x_w$ in the $\hat {\bf x}$ direction.
	
	To be more precise, we define $V_{\bf i}\equiv V_{x,y}$ the potential felt by the particles at site ${\bf i}\equiv (x,y)$. The $k^{\rm th}$ wedge-shaped obstacles is thus defined by a potential in $[0,x_w]\times [(k-1) \ell_w,k\ell_w] $:
	\begin{equation}
		V_{x,y}^{\epsilon,k} = \frac{A_{y-(k-1) \ell_w}^\epsilon}{x_{w}}(x_{w}-x)\Theta(x_{w}-x)\;,
	\end{equation}
	which is a locally linear function of $x$ with an amplitude $A^{\epsilon}_{y-(k-1)\ell_w}$ that is a linear function of $y$. Here, $\Theta(x)$ is a Heaviside step function and $\epsilon=\pm 1$ is chosen at random for each value of $k$ with equal probability, to decide the obstacles orientations. The $y$-dependent amplitudes $A^{\pm}_{y-(k-1)\ell_w}$ of the $k^{\rm th}$ obstacle is then given by:
	\begin{align}
		A^{+}_{y} =&~ \frac{\Delta V}{\ell_w} y \Theta(y) \Theta(\ell_w - y) \\
		A^{-}_{y} =&~ \frac{\Delta V}{\ell_w} ( \ell_w - y) \Theta(y) \Theta(\ell_w - y)~.
	\end{align}
	All in all, with these building blocks, the wall disordered potential $V_{x,y}$ 
	is given by
	\begin{equation}\label{eq:wallpotential}
		V_{x,y} = \sum_{k=1}^{L_y / \ell_w} V^{\epsilon_k,k}_{x,y}~,
	\end{equation}
	where $L_y$ is chosen to be an integer multiple of $\ell_w$ and $\epsilon_k$ is the orientation of the $k$-th wedge. 
	If the second wall, at $x=L_x$, is also disordered, as in Fig.~\ref{fig:nonint_two_wall}, its potential is obtained by substituting $x$ with $L_x-1-x$ in Eq.~\eqref{eq:wallpotential} and by sampling independently the orientations along the wall at $x=L_x$. This implies that the orientations of the wedges at $x=0$ and $x=L_x-1$ are independent of each other. \smallskip
	
	\textit{RTP lattice simulations:} To simulate RTPs on a square lattice, each particle is assigned an orientation $\bfu(\theta) = (\cos \theta, \sin \theta)$ where $\theta \in [0,2 \pi)$ and reorients to a new random orientation with a rate $\alpha$. In the absence of interactions and a disordered potential, the active propulsion of each particle is implemented through a biased hopping of the particles. In practice, a particle hops from a position ${\bf i}$ to any of its $2d$ nearest neighboring sites ${\bf j}$ with a rate given by $W_{{\bf i}, {\bf j}} = \max [v \bfu ({\theta}) \cdot {\bf \hat{e}}, 0]$. Here $v$ is a propulsion speed and ${\bf \hat{e}} = {\bf j} -  {\bf i}$. If ${\bf j}$ lies inside a hard wall, $W_{{\bf i}, {\bf j}}=0$. The presence of a non-zero quenched potential disorder, $V_{\bf i}$, modifies the hopping rates as $W_{{\bf i}, {\bf j}} = \max[v \bfu ({\theta}) \cdot {\bf \hat{e}} - (V_{\bf j} - V_{\bf i}),0]$. Finally, in simulations where interactions between particles are included, we take the hoping rates as $W^\mathrm{int}_{{\bf i}, {\bf j}} = W_{{\bf i}, {\bf j}} (1 - n_{\bf j} / n_M)$ where $n_{\bf j}$ is the number of particles at site ${\bf j}$ and $n_M$ is the maximal site occupancy. Such interactions are known to lead to motility-induced phase separation provided that $v/\alpha$ and the density are large enough~\cite{thompson_lattice_2011} and that $n_M>1$~\cite{soto_run-and-tumble_2014}.
	
	In what follows, we provide parameters and further details on each figure. We define the average  density as $\rho_0 \equiv N/(L_x L_y)$. \smallskip
	
	\noindent \textbf{Figure~\ref{fig:schematic}}: Instantaneous snapshots of site occupancies of passive and active particles. The passive particles are simulated with the standard Metropolis Monte-Carlo rule, with the Hamiltonian given as
	\begin{equation}
		{\cal H} = - \sum_{ \{ {\bf i}, {\bf j} \}} J n_{\bf i} n_{\bf j} + \sum_{\bf i} V_{\bf i} n_{\bf i}
	\end{equation}
	where the first summation is performed for ${\bf i}$ and ${\bf j}$ when they are the nearest neighbors of each other and each site can be occupied by at most one particle. On panel (c), we impose hard wall for $x<0$ and $L_x \leq x$ and on panel (d), we add disorder potential beside hard walls. (g) and (h) present snapshots of active particles. On panel (g), we put hard walls similarly to (c), and the tumble rates are increased by a factor of 1.5 along the walls to prevent wall accumulation. On (h), we add disordered walls along the hard walls. The parameters used are: $L_x = 2 \times 10^2$, $L_y = 5 \times 10^2$, $n_M = 1$ for (c) and (d), $n_M = 2$ for (g) and (h), $\rho_b / n_M = 0.45$, $J = 1.8 k_B T$, $v = 9.5$, $\alpha = 1$, $\Delta V = 6 k_B T$ for (c) and (d), $\Delta V = 22$ for (g) and (h), $x_w = 10$, $\ell_w = 3$, $t_0 / \Delta t = 2 \times 10^6$ for (g) and (h), and $8 \times 10^6$ Monte-Carlo sweeps have been performed before obtaining the snapshots for (c) and (d). 
	\smallskip
	
	\noindent \textbf{Figure~\ref{fig:rho_and_jy}}: The steady-state density and the current for a single realization of the disordered wall at $x=0$. On panel (a), the steady-state density $\rho(x,y)$ is defined as the time average of the number of particles $n_{\bf i}$ at ${\bf i} = (x,y)$. On panel (b),  we present the density modulation $\phi(x,y) \equiv \rho(x,y)- \langle \rho(x) \rangle$ measured with respect to $\langle \rho(x)\rangle $, defined as the time average of $ \rho(x)  \equiv L_y^{-1} \sum_{y=0}^{L_y} n_{x,y}$. 
	
	On panels (c) and (d), the current along the $y$-axis $J_y(x,y)$ is measured as follows. We define $h_{y,{\bf i}}$, the number of particles that hop from ${\bf i} = (x,y)$ to $(x,y+1)$ during a time interval $t_0$, and then evaluate the current as $J_y(x,y) = (h_{y,{\bf i}} + h_{y{\bf i-1}})/(2 t_0)$.

	The data are normalized using $\delta \phi$ and $\delta J_y$, which are the standard deviations of the density and of $J_y$ computed for each site and averaged over the whole lattice.
	
	The parameters used are: $L_x = 2^9$, $L_y = 2^9$, $\rho_0 = 0.2$, $v = 10$, $\alpha = 1$, $\Delta V = 20$, $x_w = 10$, $\ell_w = 4$, and $t_0/\Delta t = 2 \times 10^7$ with $\Delta t = (\alpha + \sqrt{2}v + 2 \Delta V)^{-1}$ the unit time of the lattice simulation.  \smallskip
	
	\noindent \textbf{Figure~\ref{fig:Single ratchet}}: The steady-state density and current streamlines for an isolated localized deformation at $x=0$. Here, the deformation is modelled by the potential
	\begin{equation} \nonumber
		V_{x,y} = \frac{\Delta V}{\ell_w} (y-x) \Theta(y-x) \Theta (\ell_w+x-y) \Theta(x_w - x)~.
	\end{equation}
	The streamlines shown in Fig.~\ref{fig:Single ratchet}(b) are obtained using the streamline plot module of OriginLab. The parameters used are: $L_x = 2.2 \times 10^2$, $L_y = 6.6 \times 10^2$, $\rho_0 = 1.0$, $v = 6$, $\alpha = 1$, $\Delta V = 10$, $x_w = 6$, $\ell_w = 5$, and $t_0 / \Delta t = 6 \times 10^7$.
	
	\smallskip
	
	\noindent {\bf Figure~\ref{fig:nonint_scaling}}: A plot of the steady-state  two-point density correlation function $\overline{\phi(x,y) \phi(x,y+\Delta y) }_c$ in the presence of a disordered wall at $x=0$. The simulation data used to evaluate the correlations are coarse-grained, so that the value of the correlation function at ${\bf i}$ is obtained by taking an average over the $5 \times 5$ lattice sites centered at ${\bf i}$. The data are then fitted to Eq.~\eqref{eq:FSC} to extract the value of $A_\rho \equiv \ell_p S_d^{-1} (2 \mu p \sigma / {\cal D}_\mathrm{eff})^2$. We then normalized the data by $A_\rho$ and added the finite-size correction ${\pi \ell_p}/{L_y}$ before comparing to the theoretical prediction, consistent with the finite-size results of Appendix~\ref{app:periodic walls}.
	
	The parameters used are: $L_x = 3 \times 10^2$, $L_y = 1.5 \times 10^3$, $\rho_b = 0.2$, $v = 10$, $\alpha = 1$, $\Delta V = 20$, $x_w = 10$, $\ell_w = 4$, $t_0 / \Delta t = 4 \times 10^6$. The disorder average is taken over $1.3 \times 10^3$ independent realizations.  \smallskip
	
	\noindent {\bf Figure~\ref{fig:current_scaling}}: Scaling of the current-current correlation function. To produce this figure, we measure the current two-point correlation function $\overline{{\bf J}(x, y) \cdot {\bf J}(x, y+ \Delta y) }_c$ with the current measured using the procedure described above, but extended to include the current in the $\hat{\bf x}$ direction. Then, a Fourier transform is carried out along the $\hat{\bf y}$ direction.
	
	The parameters used are: $L_x = 2^8$, $L_y = 2^{10}$, $\rho_b = 0.2$, $v = 10$, $\alpha = 1$, $\Delta V = 20$, $x_w = 10$, $\ell_w = 4$. $t_0 / \Delta t = 2 \times 10^6$. Disorder averages are taken over $8 \times 10^2$ independent realizations. \smallskip
	
	\noindent {\bf Figure~\ref{fig:nonint_two_wall}}: The density two-point correlation function in the presence of two disordered walls at $x=0$ and $x=L_x-1$. Similarly to the figure with a single disordered wall, we coarse-grain the data over $5 \times 5$ lattice sites. To compare the simulation data and the analytic expression~\eqref{eq:rhorho two walls}, we use the strength of the random forcing and the finite-size offset as fitting parameters. The theoretical contour lines correspond to the boundaries of the levels of the color bar.
	
	The parameters used are: $L_x = 2 \times 10^2$, $L_y = 8 \times 10^2$, $\rho_b = 0.2$, $v = 10$, $\alpha = 1$, $\Delta V = 20$, $x_w = 10$, $\ell_w = 4$. $t_0 / \Delta t = 4 \times 10^6$. Disorder averages are taken over $3 \times 10^3$ independent realizations. \smallskip
	
	\noindent {\bf Figure~\ref{fig:int_cor}}: The current two-point correlation function obtained for interacting RTPs in the presence of a disordered wall at $x=0$. The steady-state density is measured using the procedure described above. The finite-size offset of the correlation function is used as a fitting parameter. 
	The parameters used are: $L_x = 3 \times 10^2$, $L_y = 9 \times 10^2$, $n_M = 2$, $\rho_b = 0.8$, $v = 9.5$, $\alpha = 1$, $\Delta V = 18$, $x_w = 10$, $\ell_w = 3$, and $t_0 / \Delta t = 3 \times 10^5$.  Note that the tumbling rate is locally enhanced to $\alpha = 3$ when $V_{\bf i} \neq 0$ to reduce accumulation of the particles along the wall, hence enhancing the signal far away from the wall. The disorder average is taken over $5 \times 10^2$ independent realizations.
	\smallskip
	
	\noindent {\bf Figure~\ref{fig:Imry-Ma}}: The steady-state density for interacting RTPs with and without boundary disorder for a single realization of disorder. The parameters used are: $L_x = 2\times 10^2$, $L_y = 6 \times 10^2$, $n_M = 2$, $\rho_b = 0.9$, $v = 9.5$, $\alpha = 1$, $t_0 / \Delta t = 10^5$. For the panel (a), we set $\Delta V = 0$ and for the panel (b) and (c), we set $\Delta V = 18$, $x_w = 10$, $\ell_w = 3$.   \smallskip
	
	\noindent{\bf SM Movie 1:} The system is simulated in the presence of flat walls until MIPS is observed in the form of a single macroscopic liquid droplet coexisting with a gaseous background. To prevent wall accumulation, the tumble rates are increased by a factor of $1.5$ along the walls. At $t=0$, disorder along the wall is turned on and the system is let to relax. The parameters used are: $L_x = 2\times 10^2$, $L_y = 5 \times 10^2$, $n_M = 2$, $\rho_b = 0.9$, $v = 9.5$, $\alpha = 1$, $\Delta V = 22$, $x_w = 10$, $\ell_w = 3$, $t_0 / \Delta t = 2 \times 10^6$.
	
	\section{Multipole expansion for pairwise-interacting active particles}\label{app:multipole expansion interacting}
	\setcounter{equation}{0}
	
	In this appendix, we generalize the derivation of the far-field density modulations carried out for noninteracting ABPs and RTPs in section~\ref{sec:multipole expansion} to allow for pairwise interactions.
	The derivation is restricted to systems that are homogeneous and, for simplicity, it is carried in two dimensions. The generalization to higher dimensions is straightforward.
	
	We consider active particles evolving according to the dynamics
	\begin{align}\label{eq:interacting dynamics}
		\frac{d{\bf r}_i}{dt} =& v\bfu(\theta_i) - \mu \nabla \Big[V({\bf r}_i) + \sum_{j\ne i} \mathfrak{u}(|{\bf r}_i-{\bf r}_j|)\Big] \nonumber \\
		& +\sqrt{2\mathcal{D}_t} \boldsymbol{\eta}_i\left(t\right)\ , \\
		\frac{d\theta_i}{dt} =& \sqrt{2\mathcal{D}_r}\xi_i(t)\ .
	\end{align}
	In addition, the particles' orientations undergo tumbles with rate $\alpha$. Note that, in comparison to Eqs.~\eqref{eq:Langevin ABPs RTPs 1}-\eqref{eq:Langevin ABPs RTPs 2} of the main text, we now allow for interactions between the particles through a pair potential $\mathfrak{u}\left(\left|{\bf r}_i - {\bf r}_j\right| \right)$.
	
	To proceed, we use It\^o calculus~\cite{Dean:1996:JPA} to derive an equation for the empirical distribution
	$\psi\left(\bfr,\theta,t\right) = \sum_i \delta^{(2)}(\bfr_i-\bfr)\delta(\theta-\theta_i)$. From this it is  straightforward~\cite{solon_pressure_2015-3,solon_generalized_2018,granek2020bodies} to write a continuity equation for the average density field $\rho(\bfr, t)= \langle \hat{\rho}(\bfr, t) \rangle=\langle \sum_i \delta^{(2)}\left(\bfr - \bfr_i\right)\rangle$,
	\begin{equation}\label{eq:continuity interacting}
		\partial_t \rho = -\nabla\cdot {\bf J}\;.
	\end{equation}
	Here, $\bfJ$ is the average particle current and is given by
	\begin{equation}\label{eq:J int explicit}
		\bfJ = - \mu \rho \nabla V + \mu \ell_p \nabla \cdot \left[ (\nabla V) {\bf m} \right] + \mu \nabla \cdot \boldsymbol{\sigma} ~,
	\end{equation}
	where the divergence operator is contracted with $\nabla V$ in the second term and $ m_\alpha= \langle \hat{m}_\alpha(\bfr, t) \rangle = \langle \sum_i u_\alpha\left( \theta_i\right) \delta^{(2)}\left(\bfr - \bfr_i\right) \rangle$. 
	$\boldsymbol{\sigma}$ appearing in Eq.~\eqref{eq:J int explicit} can be interpreted as the stress tensor of the active fluid and is given by
	\begin{equation}
		\sigma_{ij} = -\frac{\mathcal{D}_{\rm eff}}{\mu}\rho \delta_{ij} + \sigma^{\rm P}_{ij} + \sigma^{\rm IK}_{ij}\label{eq:sigma decomposition} \;,
	\end{equation}
	with 
	\begin{align}\label{eq:sigma P}
		\boldsymbol{\sigma}^{\rm P}\left({\bf r}\right) &= \ell_p \intop d^2 r'\,\nabla \mathfrak{u}\left(\left|{\bf r}-{\bf r}'\right|\right) \langle \hat{{\bf m}} \left({\bf r}\right) \hat{\rho}\left({\bf r}'\right)\rangle \nonumber \\
		& + \frac{\mathcal{D}_t\ell_p}{\mu}\nabla{\bf m}\left({\bf r}\right) - \frac{v\ell_p}{\mu}{\bf Q}\left({\bf r}\right)\ ,
	\end{align} 
	the contribution to the stress tensor due to the local ordering of the particles' orientations. Here, ${\bf Q}=\langle \sum_i \left[ u_\alpha\left(\theta_i\right) u_\beta\left( \theta_i\right) - \frac{1}{2}\delta_{\alpha\beta} \right]\delta^{(2)}\left(\bfr - \bfr_i\right)\rangle$ is the nematic tensor, and 
	\begin{align}
		\boldsymbol{\sigma}^{\rm IK}\left({\bf r}\right) =& \frac{1}{2}\intop d^2 r'\, \frac{{\bf r}'{\bf r}'}{\left| {\bf r}' \right|} \mathfrak{u}'\left({\bf r'}\right)\times \\
		&\times\intop_0^{1} d\lambda\,\langle \hat{\rho}\left({\bf r}+\left(1+\lambda\right){\bf r}'\right)\hat{\rho}\left({\bf r}+\left(1-\lambda\right){\bf r}'\right)\rangle \ ,\nonumber\label{eq:sigma IK}
	\end{align} 
	is the Irwin-Kirkwood stress tensor. In what follows, we focus on the steady state where $\partial_t \rho =0$.
	
	Similar to Sec.~\ref{sec:multipole expansion}, we introduce $ \boldsymbol{\mathcal{J}}^\sigma = \bfJ - \mu \nabla \cdot \boldsymbol{\sigma}$, which allows us to recast Eq.~\eqref{eq:continuity interacting} in the steady state, $\nabla \cdot \bfJ=0$, as
	\begin{equation}\label{eq:Poisson's equation sigma interacting}
		\mu \partial_i \partial_j \sigma_{ij} =  -
		\partial_i \mathcal{J}^\sigma_i\;.
	\end{equation}
	Here, summation over repeated indices is implied.
	Finally, the boundary condition ensuring a vanishing current across the wall can be written as
	\begin{equation}
		J_x(0,y) = \left(\mu \partial_j \sigma_{xj}+\mathcal{J}^\sigma_x\right)\big|_{x=0}=0\ .\label{eq:boundary condition sigma interacting}
	\end{equation}
	
	In contrast to the derivation of the main text, which deals with a scalar density field, $\boldsymbol{\sigma}$ is here a tensor field, which requires more care.
	To bring Eq.~\eqref{eq:Poisson's equation sigma interacting} to the form of a Poisson equation, we perform a Helmholtz-Hodge decomposition, isolating the divergence-less part of the stress tensor
	\begin{equation}\label{eq:Helmholtz decomposition stress}
		\nabla \cdot \sigma = -\nabla \Phi + \nabla \times \boldsymbol{\Psi}\ .
	\end{equation}
	Here $\Phi$ is a scalar potential and $\boldsymbol{\Psi}\equiv\Psi\hat{z}$ is a vector potential. 
	Taking the divergence of Eq.~\eqref{eq:Helmholtz decomposition stress} along with Eq.~\eqref{eq:Poisson's equation sigma interacting}, one finds that the field $\Phi$ obeys Poisson's equation
	\begin{equation}\label{eq:Poisson interacting}
		\mu \nabla^2 \Phi =\nabla\cdot\boldsymbol{\mathcal{J}^\sigma}\ ,
	\end{equation}
	with the boundary condition
	\begin{equation}
		\mu \partial_x \Phi\bigg|_{x=0} = \left(\mathcal{J}^\sigma_x + \mu\partial_y \Psi\right)\bigg|_{x=0}\ ,\label{eq:boundary condition Phi interacting}
	\end{equation}
	similar to the equation satisfied by the density field $\rho(\bfr)$ in the derivation of Sec.~\ref{sec:multipole expansion}. From this point on, we follow the previous derivation closely.
	The solution of Eq.~\eqref{eq:Poisson interacting} reads
	\begin{align}\label{eq:intermediate interacting}
		\mu\Phi\left({\bf r}\right) =& - \intop_0^\infty dx'\,\intop_{-\infty}^\infty dy'\,G_N\left(x,y;x',y'\right) \\
		&\qquad\qquad\qquad\times\left[\nabla'\cdot{\bf \boldsymbol{\mathcal{J}^\sigma}}'+\mu(\nabla' \times \nabla') \cdot \boldsymbol\Psi'\right] \nonumber  \\
		& -\intop_{-\infty}^\infty dy'\,G_N\left(x,y;0,y'\right)\mu\partial_x'\Phi'\bigg|_{x=0} + \mu\Phi_b\;.\nonumber
	\end{align}
	In the expression for $\mu\Phi(\bfr)$, we added  $(\nabla'\times\nabla')\cdot \boldsymbol\Psi' = \left(\partial_x'\partial_y' - \partial_y'\partial_x'\right)\Psi'=0$ and $\mu\Phi_b$, a constant whose physics we identify later.
	In the far field $|x-x'|\gg \ell_p,a$, we separate the $x'$ and $y'$ components in the first integral of Eq.~\eqref{eq:intermediate interacting} using
	\begin{equation*}
		\nabla\cdot{\bf \boldsymbol{\mathcal{J}^\sigma}}+\mu (\nabla \times \nabla) \cdot \boldsymbol\Psi \!=\! \partial_x\left[ \mathcal{J}^\sigma_x + \mu\partial_y \Psi\right] + \partial_y\left[ \mathcal{J}^\sigma_y - \mu\partial_x \Psi\right],
	\end{equation*}
	and consider the first contribution. It is non-zero, to leading order, only in the vicinity of the wall and of the obstacle, and we thus Taylor-expand the Green's function in the first line of Eq.~\eqref{eq:intermediate interacting} in $x'$ near $x'=0$. As $(x')^2(\partial_x')^2 G_N \ll G_N$ in the far field, the expansion can be truncated at zeroth order to find
	\begin{align}\label{eq:Phi}
		\mu\Phi({\bf r}) &\underset{x\gg \ell_p,a}{\simeq}   \mu\Phi_b \\
		& - \intop_{0}^{\infty} dx '\intop_{-\infty}^{\infty} dy '\, G_N(x,y;x',y')\partial_y' \left[{\mathcal{J}^\sigma_y}' - \mu\partial_x' \Psi'\right]\;.\nonumber
	\end{align}
	Using the definition of $\boldsymbol{\mathcal{J}^\sigma}$ along with the expression for the current in Eq.~\eqref{eq:J int explicit}, we find that the term appearing in the integrand of Eq.~\eqref{eq:Phi} is
	\begin{align}
		\partial_y' \left[{\mathcal{J}^\sigma_y}'\right.&\left. - \mu\partial_x'\Psi'\right] =  \\
		& \!\!\!\!\!\!\!\! \mu \partial_y'\left[-\rho\partial_y' U + \ell_p \nabla'\cdot\left\{(\nabla' U) m_y'\right\} - \partial_x' \Psi' \right]\;, \nonumber
	\end{align}
	where $U$ is the part of the potential that is not invariant under translations along the ${\bf \hat{y}}$ direction.
	We then integrate Eq.~\eqref{eq:Phi} by parts. The leading order contribution in the far field comes solely from the term involving $\rho\partial_y' U$, as it contains less derivatives. With this, the scalar potential $\Phi(\bfr)$ becomes
	\begin{equation}\label{eq:multipole expansion Phi}
		\Phi\left({\bf r}\right) \underset{x\gg a,\ell_p}{\simeq} \Phi_b + \frac{1}{\pi} \frac{y p}{r^2} + \mathcal{O}({1}/{r^2})\;,
	\end{equation}
	where ${\bf p} = p{\bf \hat{y}}$ is given by the same expression as the force monopole obtained in Eq.~\eqref{eq:dipole moment} of the main text.

	To obtain the density profile, we assume that, in the far field, the stress tensor is dominated by local contributions. It is then possible to express $\boldsymbol{\sigma}$ using a gradient expansion 
	\begin{equation}
		\sigma\left({\bf r}\right) = \sigma\left(\rho\left({\bf r}\right)\right) + \mathcal{O}\left(\nabla\rho\right)\ .
	\end{equation}
	Note that the dependence of the solution on $\Psi$ does not enter explicitly into the expression~\eqref{eq:multipole expansion Phi} for $\Phi$;
	it can be shown that while $\Phi\sim\mathcal{O}\left(1/r\right)$, the contribution from $\Psi$ is of higher order, namely $\Psi\sim\mathcal{O}\left(1/r^2\right)$ (for a detailed discussion, see Ref.~\cite{granek2020bodies}). 
	This makes it possible to obtain the pressure directly from Eq.~\eqref{eq:multipole expansion Phi}, using  $P=-\frac{1}{2}{\rm Tr}\,\boldsymbol{\sigma}$ 
	\begin{equation}
		P\left({\bf r}\right) = \Phi_b + \frac{1}{\pi}\frac{yp}{r^2} + \mathcal{O}\left(1/r^2\right) \ ,
	\end{equation}
	which allows us to identify $\Phi_b = P_b$ with the bulk pressure of the system~\cite{solon_generalized_2018}. 
	
	Finally, for the steady-state density profile, we note that, to leading order in the far field, the fluid is barotropic $P\left({\bf r}\right)\approx P\left(\rho\left({\bf r}\right)\right)$. Then, neglecting higher orders in the gradient expansion, we expand the pressure near its bulk value
	\begin{equation}
		P\left({\bf r}\right) \underset{x\gg a,\ell_p}{\simeq} P_b + \left(\rho\left({\bf r}\right) - \rho_b\right) P'\left(\rho_b\right) + \mathcal{O}\left(\left(\rho - \rho_b\right)^2,\nabla\rho\right),
	\end{equation}
	with $\rho_b$ the bulk density, $\left[-\rho_b P'\left(\rho_b\right)\right]$ the inverse compressibility, and $P'=\partial P/\partial \rho$. Inverting this relation, we find
	\begin{equation}
		\rho\left({\bf r}\right) \underset{x\gg a,\ell_p}{\simeq} \rho_b + \frac{1}{\pi P'\left(\rho_b\right)}\frac{yp}{r^2} + \mathcal{O}\left(1/r^2\right)\ ,
	\end{equation}
	similar to the expression given in Eq.~\eqref{eq:multipole expansion density} for non-interacting particles.

	\section{Periodic walls and finite-size corrections}\label{app:periodic walls}
	\setcounter{equation}{0}
	
	In this appendix, we first show that periodic and aperiodic walls have very different effects on the surrounding active fluid: While aperiodic walls induce long-range modulations in the bulk of the system, periodic walls only affect a finite-size boundary layer in their vicinity.
	
	To see this in detail, we consider the effect of periodic walls on the steady-state density profile. To do so, we confine the active particles to a semi-infinite cylindrical shell $x>0$, with $y$ periodic along the $\hat{\bf y}$ direction, with a period $L_y$. By computing the Green's function in this geometry, it becomes evident that the periodicity leads to density modulations confined to a region next to the wall, of typical scale $L_y$. For simplicity, we present the derivation in two-dimensions. 
	Note that the main difference with the derivation of the main text is the different Green's function that needs to be used on the cylindrical shell. As we show, the derivation also allows us to obtain the finite-size corrections to the density-density correlation function used in the main text. This is important when comparing our theory to numerical results obtained in finite systems.

\begin{figure}
		\includegraphics[width=0.8\linewidth]{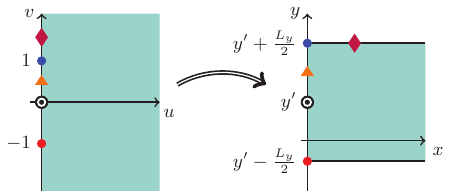}
		\caption{To compute the Green's function in Eq.~\eqref{eq:periodic Greens function}, we start from that of a point source close to an infinite wall, given in Eq.~\eqref{eq:green}. The domain is then transformed using a combination of conformal mapping techniques and properties of the Green's function into the domain shown in the right, where periodic boundary conditions are used along the $\hat {\bf y}$ direction.}\label{fig:conformal mapping domains}
	\end{figure}

	We first consider the case of an isolated deformation localized at some position $y'$.
	To obtain the correct Green's function, 
	we note that Eq.~\eqref{eq:Poisson's eq} possesses conformal invariance~\cite{kevorkian1990partial} which allows to employ a conformal mapping. Consider first the Neumann-Green's function given by Eq~\eqref{eq:green}, which corresponds to a Poisson equation with a point source located at the origin in the half-plane $(u,v)$, $u>0$ and $-\infty<v<\infty$:
	\begin{equation}
		G_N(u,v)=-\frac 1 {2\pi}\ln(u^2 + v^2)\;.
	\end{equation}
	Next, place the periodic boundary conditions of our cylindrical domain to be at $(y-y')=\pm \frac{L_y}{2}$, such that they are equally distant from the source. The mapping between the two domains is a textbook  problem on conformal mappings.
	Once complexified~\footnote{The complexified domains are also compactified, adding a ``point at infinity''~\cite{sarason1994notes}.}, it is obtained by a Schwartz-Christoffel transformation~\cite{ablowitz2003complex} as illustrated in Fig.~\ref{fig:conformal mapping domains}, which reads:
	\begin{align}
		u(x,y;y') =& \sinh\left(\frac{\pi x}{L_y}\right)\cos\left(\frac{\pi \left(y-y'\right)}{L_y}\right)\ ,\nonumber\\
		v(x,y;y') =& \cosh\left(\frac{\pi x}{L_y}\right)\sin\left(\frac{\pi \left(y-y'\right)}{L_y}\right)\ .
	\end{align}
	Let us briefly comment on why this mapping ensures the correct boundary conditions in the $(x,y)$ cylindrical domain. The points $(u=0,-1<v<1)$ are mapped onto the boundary $(x=0,y)$. The Neumann boundary conditions in the $(u,v)$ plane then ensures Neumann boundary conditions along the  $(x=0,y)$ segment. Then, the lines $(u=0,v>1)$ and $(u=0,v<-1)$ are mapped onto $(x,y=y'+\frac{L_y}{2})$ and $(x,y=y'-\frac{L_y}{2})$, respectively. Since conformal mapping preserve angles, the Neumann boundary conditions along $(u=0,|v|>1)$ ensure that the derivatives of the mapped Green's function along $\hat {\bf y}$ vanish at $(x,y'\pm\frac{L_y}2)$. Finally, the symmetry of the Green's function in the $(u,v)$ plane with respect to $v\to -v$ at $u=0$ ensures the periodicity of the mapped Green's function. 
	
	On the cylindrical shell, the Green's function of a point source localized at $(0,y')$ is thus
	\begin{align}
		G\left(x,y;0,y'\right) =& -\frac{1}{2\pi} \ln\left[u^2(x,y;y')+v^2(x,y;y')\right] \label{eq:G_N u,v} \\
		=& -\frac{1}{2\pi} \ln\left[\sinh^2\left(\frac{\pi x}{L_y}\right)\cos^2\left(\frac{\pi \left(y-y'\right)}{L_y}\right)\right.\nonumber \\
		& \left.+ \cosh^2\left(\frac{\pi x}{L_y}\right)\sin^2\left(\frac{\pi \left(y-y'\right)}{L_y}\right)\right]\ .\label{eq:periodic Greens function}
	\end{align}
	Finally, using the derivation of the main text, one then finds that a localized deformation on the wall around $(x,y) = (0,0)$ leads to a density modulation that is given in the far field by
	\begin{equation}\label{eq:Greens function dipole periodic}
		\rho(\bfr) \underset{x\gg a,\ell_p}{\simeq} \rho_b + \frac{\mu p}{\mathcal{D}_{\rm eff} L_y}\frac{\sin\left(\frac{2\pi y}{L_y}\right)}{\cosh\left(\frac{2\pi x}{L_y}\right) - \cos\left(\frac{2\pi y}{L_y}\right)}\;,
	\end{equation}
	with
	\begin{equation}
		p = -\intop_0^\infty dx'\,\intop_{0}^{L_y} dy'\,\rho\left({\bf r}'\right)\partial_y'U\;.
	\end{equation}

	We study this result in two limits. First, we note that, far from the wall, for $x\gg L_y$, the density profile decays exponentially as $\exp\left(-2\pi x/L_y\right)$. The long-range decay observed next to an infinite, aperiodic wall is thus screened due to the periodic boundary conditions and decays exponentially on a scale set by the wall's periodicity.
	
	Second, the exact expression for the Green's function obtained here allows us to understand how to compare the simulation results, obtained with a periodic disordered boundary condition along the ${\bf \hat{y}}$ direction, with the two-point correlation function computed in the main text~\eqref{eq:rhorho} for a semi-infinite system.
	To do so, we calculate the connected two-point correlation function of the density field and analyze it in the $x\ll L_y$ limit.
	In the steady state, the far-field behavior of the two-dimensional density profile generated by a disordered wall reads
	\begin{equation*}\label{eq:FSCapp}
		\rho(x,y) \!\simeq \!\rho_b + \frac{\mu}{\mathcal{D}_{\rm eff} L_y} \!\!\! \intop_{-\frac{L_y}2}^{\frac{L_y}2} \!\!\! dy'\frac{\sin\left(\frac{2 \pi (y-y')}{L_y}\right) f(y')}{\cosh\left(\frac{2 \pi x}{L_y}\right) - \cos\left(\frac{2 \pi (y-y')}{L_y}\right)}.
	\end{equation*}
	Here, $f(y)$ is the force-monopole density, satisfying $\overline{f(y)}=0$ and $\overline{f(y)f(y')}=2p^2\sigma^2\delta(y-y')$, as in Eq.~\eqref{eq:stat}. 
	With this, the disorder-averaged connected pair correlation function can be computed. For a periodicity $L_y$ much larger than the separation between the points or their distances from the wall $x,x',\Delta y\ll L_y$, one finds
	\begin{align} \label{eq:FSC}
		\overline{\rho(x,y)\rho(x',y')}_c \simeq & \frac 2{\pi} \left(\frac{\mu p \sigma}{\mathcal{D}_{\rm eff}}\right)^2 \left[\frac{(x+x')}{(x+x')^2+(y-y')^2}\right.\nonumber \\
		&\qquad\quad\left.- \frac{\pi}{L_y} + \mathcal{O}\left(1/L_y^2\right)\right]\;.
	\end{align}
	Note that to leading order in large $L_y$, this result coincides with the correlation induced by an infinite wall~\eqref{eq:rhorho}, showing a long-range decay of correlations.
	
	Finally, the fact that the results obtained in the semi-infinite domain and using periodic boundary conditions agree up to a constant that decays as $1/L_y$ is a standard property of connected correlation functions. Integrating the left-hand-side of Eq.~\eqref{eq:FSC} along $\hat {\bf y}$ or $\hat {\bf y}'$ has to vanish by definition of the connected correlation function. The Lorentzian being strictly positive, a constant has to be subtracted from the semi-infinite domain solution. In the geometry considered here, we could use our exact result for the periodic Green's function to predict this constant exactly. In the other geometries considered in the article, and in the presence of interactions, we use this offset as a fitting parameter.

	\if{
		To connect this result to finite-size corrections observed in numerical simulation, first note that 
		\begin{equation}\label{eq:truc}
			\int dy' \overline{\rho(x,y)\rho(x',y')}_c=\int dy' \overline{\rho(x,y)\rho(x',y')}-\overline{\rho(x)} \overline{\rho(x')}\;,
		\end{equation}
		where $\rho(x)=\frac{1}{L_y}\int_0^{L_y} \rho(x,y)$. Equation~\eqref{eq:truc} thus leads to
		\begin{equation}\label{eq:truc2}
			\overline{\rho(x,y)\int dy'\rho(x',y')}_c= \overline{\rho(x,y)\int dy'\rho(x',y')}-\overline{\rho(x)} \overline{\rho(x')}=0
		\end{equation}
		since $\int dy'\rho(x',y')\simeq \overline{\int dy'\rho(x',y')}$ for large systems. Clearly, the Green's function computed in semi-infinite system has a positive sign, so that it cannot satisfy Eq.~\eqref{eq:truc}. The computation leading to Eq.~\eqref{eq:FSC} shows that the periodic boundary conditions along the $\hat {\bf y}$ direction lead to a constant offset that enforces Eq.~\eqref{eq:truc2}.
	}\fi
	
	\section{Two parallel walls}\label{app:two walls}
	\setcounter{equation}{0}
	
	In this appendix, we consider the density modulation induced by two parallel disordered walls.
	The disorder on each wall acts as random forcing independent of the disorder on the  other wall. Its effect is modeled as a disordered force-monopole density, which satisfies
	\begin{eqnarray}
		&&\overline{f^\alpha_i({\bf r}_\parallel)} = 0\;,\nonumber \\
		&&\overline{f^\alpha_i({\bf r}_\parallel)f^\beta_j({\bf r}_\parallel')} = 2p^2\delta^\parallel_{ij}\sigma^2 \delta^{\alpha \beta} \delta^{(d-1)}({\bf r}_\parallel-{\bf r}_\parallel')\;,
	\end{eqnarray}
	with the indices $\alpha,\beta$ denoting either the left or the right wall. In $d=2$, the exact form of the connected, two-point correlation function can be obtained by using a conformal mapping on the one-wall solution. Here we work in arbitrary dimension and follow the simpler approach of evaluating the correlation function to leading order in the separation between the walls, $L_x$. 
	
	In this limit, we can simply sum independently the contributions of the two walls, each computed with the Green's function of the semi-infinite domain, which leads to a density profile
	\begin{align*}
		&\rho(x,{\bf r}_\parallel) \simeq \rho_b + \frac{\mu}{\mathcal{D}_{\rm eff}S_d} \intop d^{d-1}r_\parallel \nonumber \\ 
		&\left[\frac{({\bf r}_\parallel - {\bf r}_\parallel')\cdot {\bf f}^L ({\bf r}_\parallel')} {[x^2 + |{\bf r}_\parallel - {\bf r}_\parallel'|^2]^\frac{d}{2}}
		+\frac{({\bf r}_\parallel - {\bf r}_\parallel')\cdot {\bf f}^R({\bf r}_\parallel')}{[(L_x-x)^2 + |{\bf r}_\parallel - {\bf r}_\parallel|^2]^\frac{d}{2}}\right] \;.
	\end{align*}
	With this, it is straightforward to see that the connected, two-point correlation function is given by
	\begin{align}\label{eq:rhorho two walls}
		\overline{\rho(x,{\bf r}_\parallel)\rho(x',{\bf r}_\parallel')}_c \!&\approx\!\! \frac{1}{S_d}\!\!\left(\frac{\mu p \sigma}{\mathcal{D}_{\rm eff}}\right)^{\!\!2}\!\! \left[ \frac{(x+x')}{[(x+x')^2 +  |{\bf r}_\parallel - {\bf r}_\parallel'|^2]^\frac{d}{2}}\right. \nonumber \\
		& \left. + \frac{\left\{2L_x-(x+x')\right\}}{[2L_x-(x+x')]^2 +  |{\bf r}_\parallel - {\bf r}_\parallel'|^2]^\frac{d}{2}}\right].
	\end{align}
	Figure~\ref{fig:nonint_two_wall} compares this result with numerical measurements of the correlation function in two dimensions. Note that, since the numerics use periodic boundary conditions in the $\hat{\bf y}$ direction, the expression is expected to fit the data only for $\left|y - y'\right| \lesssim L_y$. To account for the finite size of the simulation box, following the results of Appendix~\ref{app:periodic walls}, we include a constant offset when we fit our numerical data to Eq.~\eqref{eq:rhorho two walls}.

	\section{Structure factor of the linear field theory} \label{app:structure factor}
	\setcounter{equation}{0}
	
	Here we detail the calculation of the structure factor of our linear field theory. We start by writing the equation for the density modulation $\phi({\bf r}, t)$ using Eqs.~\eqref{eq:field} and~\eqref{eq:J}. Performing a Fourier transform gives 
	\begin{equation} \label{eq:app_fourier_field_linear}
		\partial_t \phi({\bf q}, t) = - q^2 g[\phi] - i {\bf q} \cdot \left[  {\bf f}({\bf q}) + \sqrt{2D} \boldsymbol{\eta} ({\bf q},t) \right],
	\end{equation}
	where $g[\phi]=u\phi + K  q^2 \phi$ is the Fourier transform of the chemical potential.
	The statistics of ${\bf f}({\bf q})$ and $\boldsymbol{\eta} ({\bf q},t)$ are obtained using Eqs.~\eqref{eq:eta} and~\eqref{eq:fstat} to give
	\begin{align}
		\nonumber \overline{f_i(q_x, {\bf q}_\parallel)}=&0\;, \\
		\nonumber \overline{f_i(q_x, {\bf q}_\parallel) f_j(q_x', {\bf q}_\parallel') } =&\, 2  s^2 \delta^\parallel_{ij} (2 \pi)^{d-1} \delta^{(d-1)} ({\bf q}_\parallel + {\bf q}_\parallel)\;, \\
		\nonumber \langle{\eta_i({\bf q},t)}\rangle=&0\;, \\
		\nonumber \langle{\eta_i ({\bf q}, t) \eta_j({\bf q}', t')}\rangle =&\,  \delta_{ij} (2 \pi)^d \delta^d ({\bf q} + {\bf q}') \delta (t - t')\;,
	\end{align}
	Here, we decompose the wave vector as ${\bf q} = (q_x, {\bf q}_\parallel)$ where $q_x$ is its $x$-component and ${\bf q}_\parallel$ corresponds to the components parallel to the wall. 
	
	The structure factor can then be computed directly by solving the linear inhomogeneous differential equation for $\phi({\bf q},t)$ and evaluating $\overline{\langle \phi({\bf q}) \phi({\bf q'}) \rangle}$, leading to Eq.~\eqref{eq:structure_factor}. 
	The large-scale asymptotic expression of the structure factor is obtained by taking the limit of small $q$. Performing an inverse Fourier transform on Eq.~\eqref{eq:structure_asymptotic} then shows that the structure factor is consistent with the two-point correlation function predicted by the multipole expansion of Eq.~\eqref{eq:rhorho}.
	
	\section{The Imry-Ma argument for a droplet configuration} \label{app:Imry-Ma}
	\setcounter{equation}{0}
	
	\begin{figure}
		\includegraphics[width=0.8\linewidth]{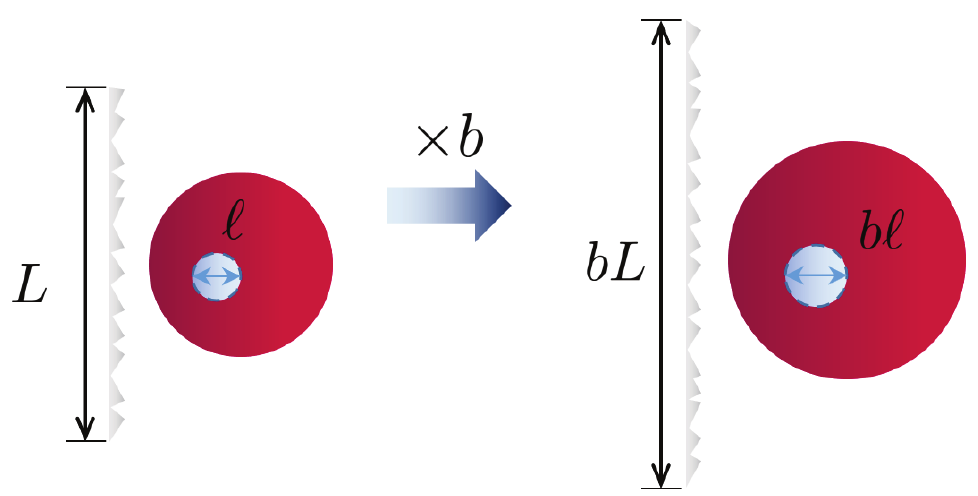}
		\caption{The scaling argument presented in Fig.~\ref{fig:Imry-Ma} for the wetting layer can be adapted to the case of a liquid droplet. Upon scaling the system size by a factor of $b$, we compare the scaling of the bulk and surface energies, whose competition determines if the droplet is macroscopically stable. }
		\label{fig:app_droplet}
	\end{figure}
	
	Here we present an alternative version of the Imry-Ma argument presented in Sec.~\ref{sec:Imry-Ma}. In contrast to the case discussed in the main text, we do not assume that the phase-separated state wets the wall. Instead, we consider a liquid bubble in the bulk of the system, surrounded by a gaseous phase, following the seminal work of Ref.~\cite{imry_random-field_1975}. As we show, both approaches lead to similar conclusions.
	
	We consider the droplet of linear size $\ell$ shown in Fig.~\ref{fig:app_droplet}. 
	The surface contribution to the free-energy scales as $E_\gamma = \gamma \ell^{d-1}$. To check the stability of this configuration, this should be compared to the bulk energy contribution of the effective potential induced by the boundary disorder $E_U = \int_{\ell^d} \mathrm{d}^d {\bf r}~ \rho_0 U({\bf r})$, where the integral goes over the volume of the droplet, which scales as $\ell^d$. 
	We now compare the scaling behaviors of $E_\gamma$ and $|E_U|$ under an increase of the system size by a factor of $b$, ${\bf r} \to b {\bf r}$. The surface energy scales as $E_\gamma \to b^{d-1} E_\gamma$, while the  scaling of the $|E_U|$ can be estimated from
	\begin{align}
		\nonumber \overline{ E_U^2 } &\to \rho_0^2 \int_{(b\ell)^d} \mathrm{d}^d {\bf r} \int_{(b\ell)^d} \mathrm{d}^d {\bf r}' ~  \overline{ U({\bf r}) U({\bf r}')} \\
		\nonumber &=  b^{d+1} \int_{\ell^d}  \mathrm{d}^d {\bf r} \int_{\ell^d} \mathrm{d}^d {\bf r}' ~ \frac{ (s^2 \rho_0^2/S_d) (x + x')}{ \left[ (x + x')^2 + | \Delta {\bf r}_\parallel|^2 \right]^{d/2} }~, 
	\end{align}
	where the change of variables ${\bf r} \to b {\bf r}$ was carried out in the second line. The droplet bulk energy is thus rescaled as $|E_U| \to b^{\frac{d+1}{2}} |E_U|$.
	
	Comparing the scaling of $E_\gamma$ and $E_U$ shows that the surface energy dominates the bulk contribution when the dimension satisfies
	\begin{equation}
		\frac{d+1}{2} < d-1 \quad \textit{i.e.} \quad 3 < d~.
	\end{equation}
	The analysis, as expected, agrees with the Imry-Ma argument of the main text, showing that the phase-separated state is unstable in dimensions smaller than $d_c=3$.

	\bibliography{biblio}
	
\end{document}